\documentclass[useAMS,usenatbib,usegraphicx]{mn2e}
\usepackage{graphics,amssymb,jnlabr}

\newcommand{\oiiw}{\mbox{[O II] $\lambda$3727}}

\newcommand{\oiiibw}{\mbox{[O III] $\lambda$5007}}

\newcommand{\oii}{\mbox{[O II]}}

\newcommand{\hal}{\mbox{H$\alpha$}}

\newcommand{\hb}{\mbox{H$\beta$}}

\newcommand{\hd}{\mbox{H$\delta$}}

\newcommand{\oiii}{\mbox{[O III]}}

\begin{document}

\title[Environments of K+A Galaxies]{The DEEP2 Galaxy Redshift Survey: Environments of Poststarburst Galaxies at $z\sim0.1$ and $z\sim 0.8$}
\author[R. Yan et al.]{Renbin Yan$^{1}$\thanks{E-mail:
yan@astro.utoronto.ca},
Jeffrey A. Newman$^{2}$, S. M. Faber$^{3}$, Alison L. Coil$^{4,5,6}$, 
\newauthor
Michael C. Cooper$^{4,7}$, Marc Davis$^{8,9}$, Benjamin J. Weiner$^{4}$, Brian F. Gerke$^{10}$, 
\newauthor
David C. Koo$^{3}$ \\
$^1$ Department of Astronomy and Astrophysics, University of Toronto, 50 St. George Street, Toronto, ON M5S 3H4, Canada \\
$^2$ Department of Physics and Astronomy, University of Pittsburgh, 401-C Allen Hall, 3941 O'Hara Street, Pittsburgh, PA 15620, USA \\
$^3$ UCO/Lick Observatory, Department of Astronomy and Astrophysics, University of California, Santa Cruz, CA 95064, USA \\
$^4$ Steward Observatory, University of Arizona, 933 N. Cherry Avenue, Tucson, AZ 85721, USA \\
$^5$ Hubble Fellow \\
$^6$ Department of Physics and Center for Astrophysics and Space Sciences, University of California, San Diego, CA 92093\\
$^7$ Spitzer Fellow \\
$^8$ Department of Astronomy, University of California, Berkeley, CA 94720, USA \\
$^9$ Department of Physics, University of California, Berkeley, CA 94720, USA \\
$^{10}$ Kavli Institute for Particle Astrophysics and Cosmology, Stanford Linear Accelerator Center, 2575 Sand Hill Rd., M/S 29, \\
Menlo Park, CA 94025, USA 
}

\maketitle

\begin{abstract}
Poststarburst (also known as K+A) galaxies exhibit spectroscopic signatures indicating that their star formation was recently quenched; they are candidates for galaxies in transition from a star-forming phase to a passively-evolving phase. We have spectroscopically identified large samples of poststarburst galaxies both in the Sloan Digital Sky Survey (SDSS) at $z\sim0.1$ and in the DEEP2 Galaxy Redshift Survey at $z\sim0.8$, using a uniform and robust selection method based on a cut in \hb\ line emission rather than the more problematic \oiiw. 
Based on measurements of the overdensity of galaxies around each object, we find that poststarburst galaxies brighter than $0.4L_B^*$ at low redshift have a similar, statistically-indistinguishable environment distribution as blue galaxies, preferring underdense environments, but dramatically different from that of red galaxies. However, at higher-$z$, the environment distribution of poststarburst galaxies is more similar to red galaxies than to blue galaxies. We conclude that the quenching of star formation and the build-up of the red sequence through the K+A phase is happening in relatively overdense environments at $z\sim1$ but in relatively underdense environments at $z\sim0$. Although the relative environments where quenching occurs are decreasing with time, the corresponding absolute environment may have stayed the same along with the quenching mechanisms, because the mean absolute environments of all galaxies has to grow with time. In addition, we do not find any significant dependence on luminosity in the environment distribution of K+As. The existence of a large K+A population in the field at both redshifts indicates that cluster-specific mechanisms cannot be the dominant route by which these galaxies are formed.

Our work also demonstrates that studying poststarburst-environment relations by measuring the K+A fraction in different environments, as is common practice, is highly nonrobust; modest changes in the comparison population used to define the fraction can drastically alter conclusions.  Statistical comparisons of the overall environment distributions of different populations are much better behaved.

\end{abstract}


\section{Introduction}

How bulge-dominated (early-type) galaxies form is a long-standing puzzle in galaxy formation and evolution theories. These galaxies have old stellar populations and no ongoing star formation. However, they did not all form at high redshift: many emerged at relatively recent epochs, as evidenced by the double-to-quadruple increase in their comoving number density since $z\sim1$ \citep{BellWM04, Willmer06, Faber07,Brown07}. It is poorly understood what processes shut off the star formation in their progenitors and possibly transformed their morphologies and dynamical structures. 

\begin{figure*}
\begin{center}
\includegraphics[angle=90,totalheight=0.4\textheight,viewport=0 0 420 755,clip]{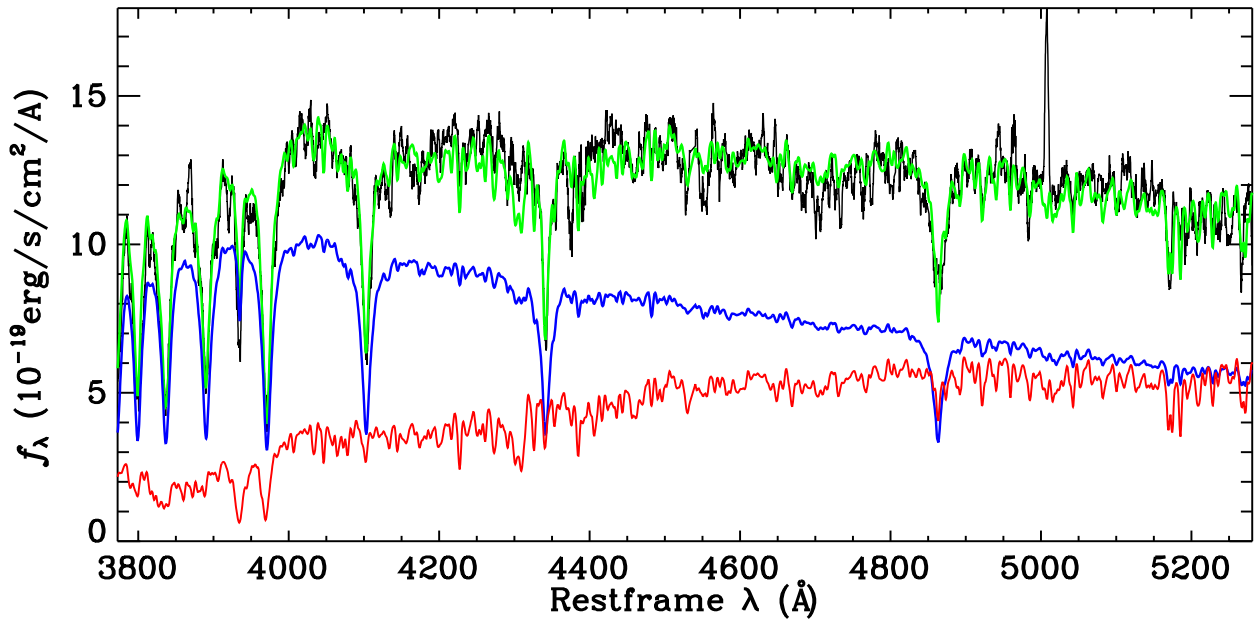}
\caption[Sample poststarburst spectrum]{Sample poststarburst galaxy spectrum taken from the DEEP2 galaxy redshift survey. The black curve shows the observed spectrum with a smoothing window of 11 pixels. The red and blue curves show the spectral decomposition. The templates used are the same as in \cite{Yan06}. The blue curve is the young stellar population component, which is dominated by A stars. The red curve is the old stellar population component, which is dominated by K--giant stars. The green curve is the linear combination of the red and blue curves and it follows the data closely. This demonstrates that the K+A model fits the spectrum fairly well. Note that the \oiiibw\ emission is prominent in this spectrum. The high \oiii/\hb\ ratio indicates this poststarburst galaxy is also a narrow-line Seyfert galaxy.
} 
\label{fig:kpa_sample}
\end{center}
\end{figure*}

Poststarburst galaxies, first identified by \cite{DresslerG83}, have been hypothesized to be the direct progenitors of early-type galaxies, since their spectra show signs of recent quenching of star formation: they possess large populations of young stars but no ongoing star formation. These galaxies are also called K+A or E+A galaxies, as their spectra can be roughly decomposed into a combination of a K giant star (or early-type galaxy) spectrum and an A star spectrum (see Fig.~\ref{fig:kpa_sample}). The visual morphology \citep{BlakePC04, Balogh05, Yang08}, S\'{e}rsic index distribution \citep{Quintero04}, kinematics \citep{Norton01}, and the elemental abundances \citep{Goto07} of poststarbursts also suggest that they are likely progenitors of early-type galaxies. Studies of poststarbursts aim to reveal how star formation is quenched and shed light on the formation of bulge-dominated galaxies. 


An important hint for solving these puzzles comes from galaxy environment studies. The environment 
distribution of poststarbursts can differentiate among possible quenching mechanisms, as in some models the quenching is caused, or partly induced, by the external environment of a galaxy (e.g., by mechanisms that are strong only in groups or clusters of galaxies). 
Comparing the environment distribution of poststarbursts to their potential progenitors, blue galaxies, and possible descendants, red galaxies, will reveal what the preferred environment for quenching is. 




Previous studies of the environment distribution of poststarbursts have found contradictory results. \cite{Dressler99}, \cite{Poggianti99}, and \cite{TranFI03,TranFI04} used samples around intermediate redshift clusters (mostly $0.3 < z< 0.6$) and found a higher poststarburst fraction in clusters than in the field; \cite{Balogh99} did a similar study around X-ray luminous clusters from CNOC1 at $0.18<z<0.55$ but found the poststarburst fraction in clusters is similar to the field; while \cite{Zabludoff96} (LCRS), \cite{BlakePC04} (2dF), \cite{Quintero04}, \cite{Goto05}, \cite{Balogh05}, and \cite{Hogg06} (SDSS in the last four) used local samples (all with $z\sim0.1$) and found poststarbursts are more likely found in the field, outside clusters. These past studies, except those using 2dF or SDSS, usually were limited in sample size, especially at higher redshifts. The methods for selecting K+A galaxies in these papers also differed both in the spectral features adopted and the threshold used. The poststarburst samples in most of these studies, except for \cite{Quintero04} and \cite{Hogg06}, are often incomplete or heavily contaminated due to selection on \oii\ equivalent width, causing the large population of poststarburst AGN hosts to be omitted, as explained in \cite{Yan06} and in Sec.~\ref{sec:select} of this paper. Therefore, the poststarburst environment distribution is poorly known at high redshift $(z>0.2)$. However, high redshift is particularly important in understanding the formation of poststarburst galaxies since their fraction among all galaxies is significantly higher at $z\sim1$ than $z\sim0$ \citep{LeBorgne06, Wild09}. Whether the environments of poststarbursts evolve with redshift is also unknown.

Most of these previous studies compared the fraction of poststarbursts among cluster galaxies with the fraction among a field sample. The result from this method depends sensitively on the parent sample in which the fraction is measured. Because galaxy properties differ significantly across different environments, the difference between poststarburst fraction across different environments can be largely due to the different mix of galaxies in the sample. A meaningful comparison is to count poststarburst galaxies relative to their potential progenitors. A varying fraction across different environments for poststarburst relative to their potential progenitors would reveal how environment affects the possibility for quenching. 

In this work, we select poststarbursts uniformly from two large redshift surveys: SDSS at $z\sim0.1$ and DEEP2 at $z>0.7$, using improved selection methods. The sample from DEEP2 is the largest clean poststarburst sample to date at high $z$. With these samples, we investigate the environment distribution of poststarbursts at both low $z$ and high $z$, and compare with their potential progenitors and possible descendants.



This paper is organized as follows. In Section 2, we describe the data used. In Section 3, we describe the selection of poststarburst galaxies, the parent sample construction, the fraction and contamination estimates, and the environment measurements. We then present the colour-magnitude distribution of poststarburst galaxies in Section 4. In Section 5, we show the environment distributions of K+A galaxies in SDSS and DEEP2, compared with blue and red galaxies. We compare our results with previous studies and discuss the evolution with redshift. In Section 6, we investigate the luminosity dependence in the K+A environment distribution. We discuss the redshift evolution of K+A environment distribution in Section 7 and the implications of these results for different quenching mechanisms in Section 8. 
We conclude in Section 9. 

Throughout this paper, we use a $\Lambda$CDM cosmology with $\Omega_{\rm M}$, $\Omega_{\rm \Lambda}$=(0.3,0.7), and a Hubble constant of $H=100{\rm h~km~s^{-1}~Mpc^{-1}}$. All magnitudes within this paper are on the AB system \citep{OkeG83}.

\section{Data}
The DEEP2 Galaxy Redshift Survey \citep[Faber et al. in prep]{Davis03, Davis07} was a three year program using the DEIMOS spectrograph on the Keck II telescope covering $\sim3$ sq. deg of sky in four widely-separated fields, to a limiting magnitude of $R_{\rm AB}=24.1$. The photometry was obtained with the CFH12K camera on the Canada-France-Hawaii Telescope \citep[see][]{CoilNK04}. To focus the spectroscopic observations on high-$z$ galaxies, in three of the four fields, a colour selection is applied to select galaxies at $z>0.75$. Tests done on the fourth field prove this pre-selection is successful: only $\sim10\%$ of the final sample is from $z<0.75$, while $<3\%$ of actual objects at $z>0.75$ are discarded. Each CFHT pointing is adaptively tiled with 40 slitmasks in an overlapping chevron pattern, giving each galaxy two chances to be on a mask. Tests done with mock catalogs indicate that dense regions are only slightly undersampled \citep{Gerke05}. Overall ~60\% of galaxies that meet the selection criteria are targeted.

We have obtained spectra for $\sim50,000$ galaxies, with a resolution of $R\sim5000$ covering the wavelength range from $6500{\rm \AA}$ to $9100{\rm \AA}$. The data are reduced by a dedicated reduction pipeline (Cooper et al. in prep). Redshifts are measured by the pipeline and confirmed by eye; $>33,000$ galaxies have yielded high-confidence redshifts ($>95\%$ confidence). 
Absolute $B$ magnitudes and restframe $U-B$ colour are computed using the K-correction code described in \cite{Willmer06}.

The SDSS \citep{York00, Stoughton02} is an imaging and spectroscopic survey that covered $\sim \pi$ steradian of the celestial sphere, utilizing a dedicated 2.5-m telescope at Apache Point Observatory. The imaging are collected with five broadband filters in drift scan mode \citep[$u,g,r,i, {\rm and}~z$;][]{Fukugita96,Stoughton02}. Spectra are obtained with two fiber-fed spectrographs, covering the wavelength range of 3800-9200\AA\ with a resolution of R $\sim2000$. The SDSS fibers have a fixed aperture of 3".

The SDSS spectroscopic data used here have been reduced through the Princeton
spectroscopic reduction pipeline (Schlegel et al.\ in prep),
which produces the flux- and wavelength-calibrated
spectra.\footnote{http://spectro.princeton.edu/} The redshift catalog of
galaxies used is from the NYU Value Added Galaxy Catalog (DR4)
\footnote{http://wassup.physics.nyu.edu/vagc/} \citep{BlantonSS05}.
K-corrections for SDSS were derived using \cite{BlantonBC03}'s {\it kcorrect}
code v3\_2. 

\section{Methods}

\subsection{Poststarburst Selection} \label{sec:select}

In principle, poststarburst galaxies are defined to have no ongoing star formation but to have experienced a significant star-forming epoch in the recent past.\footnote{An important caveat regarding the necessity of a burst will be discussed further at the end of this section} Practically, the two criteria have to be measured from specific spectral signatures, which unavoidably brings along the potential for bias, contamination, and a mixture of different physical processes.  

The first criterion, the lack of ongoing star formation, is usually indicated by the lack of line emission.
However, line emission can also result from AGN (Active Galactic Nuclei) or LINER(Low-Ionization Nuclear Emission-line Region)\footnote{It is still uncertain whether LINERs are associated with AGN. There are other non-AGN mechanisms that could produce LINER-like emission line ratios, none of which is tied to star formation. See \cite{Ho04} and \cite{Yan06} for further discussion.} activity. Therefore, a selection based on the lack of line emission will not be complete, as it will be biased against poststarburst galaxies with strong AGN/LINER-like emission. 
The level of incompleteness caused by such a selection method highly depends on the choice of emission line used. The two commonly used star-formation indicators in the optical are \oiiw\ and \hal. Most previous K+A samples employed \oii, which is more easily accessible observationally at $z>0.4$. However, as demonstrated by \cite{Yan06}, significant \oii\ emission associated with an AGN/LINER is observed in nearly half of the red galaxies and $>70\%$ of K+A galaxies (as defined using \hal), causing a large incompleteness in K+A selection. In contrast, \hal\ emission in LINERs is much weaker; they exhibit a very high \oii/\hal\ ratio. 
In addition, \oii\ emission from star-forming galaxies is much weaker than \hal\ due to dust extinction. The combined effect makes poststarburst galaxies totally indistinguishable from star-forming galaxies according to \oii\ emission, but clearly distinguishable according to \hal\ emission. Therefore, K+A selection is much cleaner if \hal\ is used as the star formation indicator.  
At $z>0.4$, we have a dilemma as \oii\ is so undesirable, while \hal\ is redshifted into the infrared. We therefore rely on \hb\ emission as a star formation indicator in this work; it is observable in optical spectra for $z<0.9$. 
Although it is much weaker than \hal, it still provides a much more complete and more robust selection of K+As than \oii\ \citep{Yan06}.


The second criterion for a galaxy to be a poststarburst, significant recent star formation, can be indicated by the presence of a large population of A stars, due to their short lifetime ($<1$Gyr). These stars will cause strong Balmer absorption features to be detectable in a galaxy's spectrum. Most past studies used the equivalent width (EW) of \hd\ in absorption \citep{Dressler99, Poggianti99, Balogh99, Goto03}, or a combination of two or three Balmer absorption lines \citep{Zabludoff96, TranFI03, TranFI04, BlakePC04}, to indicate the relative level of recent star formation. Recently, \cite{Quintero04} and \cite{Yan06} decomposed each galaxy spectrum into two components -- a young component mimicking A stars and an old component similar to K stars -- and used their ratio (dubbed A/K) to indicate the relative strength of the recent burst. We take the latter approach. However, for robustness, instead of taking the ratio between the amplitudes of the two components, we use the flux ratio between the young component and the total stellar continuum around 4500\AA, denoted by $f_{\rm A}$ for "A-star fraction," to indicate the relative level of recent star formation. This has several advantages over both EW indicators and the A/K ratio: it effectively averages over multiple Balmer absorption features, providing a less noisy measurement than the average of two or three EWs; it is more stable than the A/K ratio as the latter becomes noisy when the K component is weak; and it is less sensitive to the overall continuum shape or flux calibration as a broadly-smoothed continuum is removed before fitting. For details of the fitting procedure, see \cite{Yan06}.

\begin{figure*}
\begin{center}
\includegraphics[angle=90,totalheight=0.35\textheight,viewport=0 -10 420 730,clip]{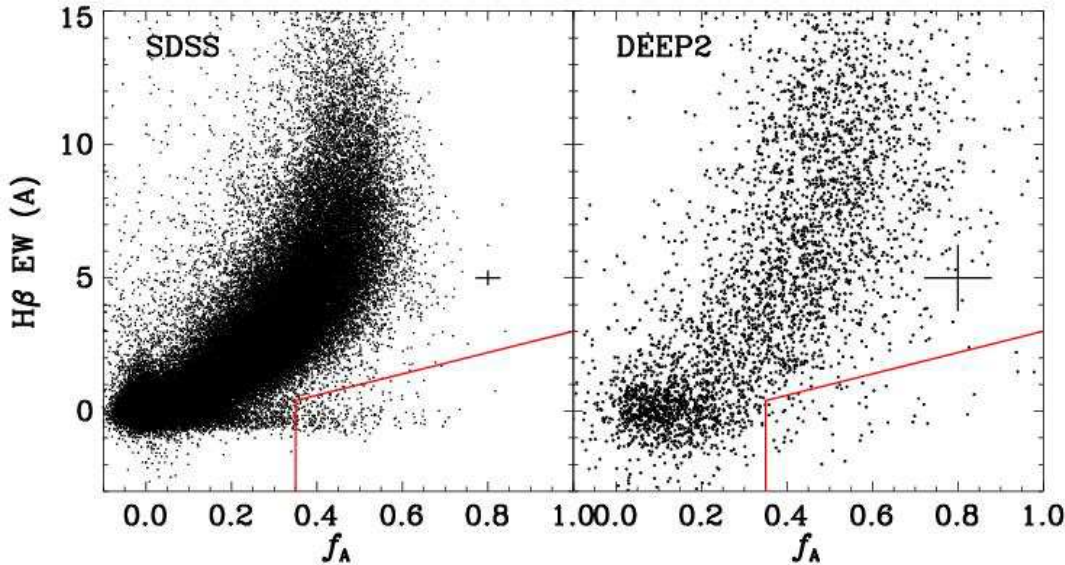}
\caption[\hb\ EW vs. $f_{\rm A}$ for SDSS and DEEP2 samples]{\hb\ EW vs. $f_{\rm A}$ for the SDSS sample (left) and the DEEP2 sample (right) as defined in Table~\ref{tab:sample}. The solid lines indicate the selection criteria adopted for identifying K+A galaxies. The crosses at position [0.8,5] indicate the median 1-$\sigma$ measurement error for $f_{\rm A}$ and \hb\ EW in each sample. In the left panel, the K+A galaxies comprise a clearly distinct population and stand out as the horizontal spur with nearly zero \hb\ EW but relatively large A-star fraction ($f_{\rm A}$). The spur is not as obvious in the DEEP2 sample due to the greater noise in spectral index measurements and smaller sample. However, as described in \S\ref{sec:contami}, we can reliably measure the contamination due to scatter in both dimensions and quantify the true K+A fraction.  
}
\label{fig:deep_1p}
\end{center}
\end{figure*}

We therefore measure the two indicators, \hb\ EW (in emission) and $f_{\rm A}$, in each galaxy spectrum from DEEP2 and SDSS. K+A galaxies are identifiable in a plot of \hb\ EW\footnote{Throughout this paper, we use positive values for emission line EW and negative values for absorption line EW.} vs. $f_{\rm A}$. As shown in Fig.~\ref{fig:deep_1p}, most galaxies exhibit a tight correlation between these parameters. This is to be expected, as galaxies with relatively strong recent star formation generally have a high ongoing star formation rate. K+A galaxies stand out as a horizontal spur in this diagram, as they have no ongoing star formation but significant recent star formation. This separation between K+A galaxies and star-forming galaxies was first demonstrated by \cite{Quintero04}. Here we use an improved recent star formation indicator, $f_{\rm A}$, in place of the A/K ratio used in \cite{Quintero04}, for reasons mentioned above. This makes the separation much clearer, even though \hb\ is weaker and harder to measure than \hal. The solid lines in the plots indicate our selection criteria for K+A galaxies, which can be expressed as the two inequalities:
\begin{eqnarray}
f_{\rm A} &>&  0.35 \\\label{eqn:fa}\nonumber 
&{\rm and}& \\ 
\hb\ {\rm EW (\AA}) &<& 4\times f_{\rm A} -1 {\rm .} \label{eqn:hb}
\end{eqnarray} 

The choice of the $f_{\rm A}$ threshold has its arbitrariness since the horizontal spur in Fig.~\ref{fig:deep_1p} is a continuous sequence from strong K+A galaxies extending to quiescent, old galaxies at (0,0). Ideally, we would put the threshold at the point above which the K+A sequence (the horizontal spur) is separable from the star-forming sequence. However, this depends on the intrinsic scatter and measurement error of each population. In the SDSS panel of Fig.~\ref{fig:deep_1p}, the separation can be achieved at least down to $f_{\rm A}$ of 0.25. But in the DEEP2 sample, due to larger measurement errors, we are forced to use a higher threshold to keep the sample clean. Therefore, we choose $f_{\rm A} > 0.35$ as the criterion for both DEEP2 and SDSS samples in order to make fair comparisons. The \hb\ cut is chosen to go through the valley between the two sequences in SDSS. Our results are less sensitive to variation in this cut as the valley is underpopulated.

An important caveat in the definition is that even strong Balmer absorption features, such as $\hd\ {\rm EW (in absorption)} < -5{\rm \AA}$, or equivalently $f_{\rm A}> 0.4$, do not guarantee the necessity of a starburst, or a significantly elevated SFR, before quenching. A sharp quenching of star formation {\it without a burst} is also possible to produce a K+A galaxy. Using \cite{BC03} population synthesis models, we found that a $5\times10^{10}{\rm M_\odot}$ (stellar mass) galaxy with a 10${\rm M}_\odot/{\rm yr}$ constant SFR would have an $f_{\rm A}$ of 0.6 (\hd\ EW$\sim -7$\AA) after abrupt quenching, which does not drop below 0.3 for another 0.5Gyr. Such a level of star formation is common at redshift $z\sim0.7$ and above \citep{NoeskeFW07}. Therefore, a starburst is not absolutely necessary for a galaxy to be classified as a K+A galaxy. This is also shown by \cite{LeBorgne06} using PEGASE-HR stellar population synthesis code \citep{LeBorgne04}.
Although the optical and NIR colours of K+A galaxies suggest enhanced recent star formation is necessary \citep{Balogh05}, the accuracy of the name 'poststarburst' is questionable. The name ``K+A'' is still precise since it only refers to the observational definition, but it lacks a reference to the physical processes. A more appropriate name for these galaxies would be ``post-quenching galaxies''. However, for consistency with the literature, we have been refering to all K+A galaxies as ``poststarburst'' galaxies. From now on, we will adopt the name ``K+A galaxy'' and ``post-quenching galaxy''.


\subsection{Parent Sample Definition}\label{sec:sample}

\subsubsection{DEEP2}

The DEEP2 sample we use is a subset of the spectra obtained by the survey. We require galaxies to have \hb\ covered in the spectrum and to not be severely impaired by sky-subtraction residuals. The spectra are also required to reach a minimum restframe wavelength below 3900\AA\ to make reliable spectral decomposition possible. This limits us to the redshift range $0.65<z< 0.88$, yielding a sample of 9564 galaxies. 
However, we cannot use them all for the following environment analysis, for a few reasons.  

First, the DEEP2 sampling density drops precipitously below $z\sim0.7$ due to the colour preselection of the survey, leading to undesirably large correction factors and high uncertainty in environment measurements, we limit the sample to $z>0.72$. This removes 14\% of the sample.

Secondly, the environment measurement is not reliable when a galaxy is too close to a survey edge. As a result, we further limit the sample to only galaxies that are at least $1h^{-1}$Mpc (comoving) away from a survey edge (see \S\ref{sec:env} and \citealt{Cooper06}). This criterion removes 23\% of the sample after the above cut. In total, we have $(1-0.14)\times(1-0.23)=66\%$ of the original sample left.


Thirdly, we select a luminosity-limited sample for the analysis. DEEP2's R-band magnitude limit ($R_{AB} < 24.1$) corresponds to a color- and redshift-dependent limit in restframe $B$-band (see \citealt{Willmer06}). This cutoff is a function of color, shallower for redder galaxies \citep{Willmer06}. At $z=0.88$, the high redshift end of the K+A window, the limit is $M_B-5\log h \sim -19.7$ for typical red galaxies with restframe $U-B=1.2$. Therefore, DEEP2 is effectively volume-limited for galaxies of all colours to $M_B-5\log h < -19.7$ for $z<0.88$. With this magnitude cut, 53\% of the sample after above cuts is further removed, most of them relatively blue galaxies. In total, we have 31\% of the original sample left. 
We also make a bright subsample using a brighter luminosity cut at $M_B-5\log h < -20.7$, which is approximately $L_*$ at $z=0.8$, to investigate the possibility of luminosity dependence. 

In principle, to make an unbiased comparison between post-quenching galaxies and their potential progenitors and descendants, it is most ideal to make a stellar-mass limited sample. However, we do not have the luxury of doing this. Although $K$-band data is available in 1/3 of the DEEP2 fields \citep{Bundy06}, the combined depth of the DEEP2 spectroscopy and the $K$-band survey is not deep enough to cover enough post-quenching galaxies to allow significant detection of any environment trend. Therefore, we resort to a $B$-band luminosity-limited sample. Our $M_B-5\log h < -19.7$ limit corresponds approximately to a stellar mass limit of $M_* > 10^{10.7} M\odot$ for red galaxies and $M_* > 10^{10.1}M\odot$ for blue galaxies. For K+A galaxies, it is approximately $M_* > 10^{10.6} M\odot$.

Lastly, we limit our sample to only galaxies with a high signal-to-noise (S/N) spectrum. When the S/N per pixel in the continuum falls below 1 (corresponding to S/N$\sim2.3$ per FWHM), the spectral decomposition becomes unstable, which also introduces large errors on the \hb\ EW. This will contaminate the K+A sample with blue, star-forming galaxies. We therefore remove all galaxies with S/N per pixel less than 1.0. This criterion removes 10\% of the sample, most of which are blue galaxies. In addition, we remove a handful (0.5\%) of galaxies which have \hb\ EW error greater than 5\AA\ or have \hb\ EW less than -5\AA (where we define emission as positive EW). These criteria effectively clean the K+A sample without changing the whole sample significantly. They only brighten the median $M_B$ by 0.06 for the whole sample, 0.02 for red galaxies, and 0.06 for blue galaxies, all of which are insignificant. In total, we have 28\% of the original sample left. Table~\ref{tab:sample} summarizes the details of our sample definition.

With the above selection, we are left with 2649 galaxies from the DEEP2 survey. This is the primary DEEP2 sample used in the following analysis. There are 39 K+A galaxies in this sample. After inspecting their spectra individually, we manually removed 5 of them which have \hb\ severely underestimated due to strong residuals from sky subtraction. The remaining 34 K+A galaxies are all robustly detected, brighter than our luminosity cut, and have well-measured environment. Whenever possible, we lift some of the selection cuts and use a larger sample. For example, in Fig.~\ref{fig:cmd}, we lift the redshift cuts, edge distance cut, and the magnitude cut, but keep the signal-to-noise cut, giving the maximum sample of 74 K+A galaxies. All of the 74 K+As are plotted in Fig.~\ref{fig:cmd}. This is the largest K+A sample ever found in this redshift range.

\subsubsection{SDSS}

To test the evolution in the K+A population from $z\sim1$ to $z\sim0$, we build a comparison dataset from the SDSS main galaxy sample in the redshift range $0.07<z<0.12$. To make a fair comparison, we select a sample down to the same $B$-band luminosity limit relative to $L_*$, the characteristic luminosity of galaxies in a Schechter-function fit, as at $z\sim0.8$. As star formation rates decline and stellar populations age, galaxies fade; $L_*$ in the $B$-band decreases by 1.3 mag per unit $z$ \citep{Willmer06, Faber07}. 
Our $B$-band limit at $z\sim0.79$ corresponds to $M_B=-18.8$ at $z\sim0.1$. However, the depth of SDSS in $B$-band is shallower than this at $z>0.082$. Therefore, we are unable to construct a volume-limited sample from SDSS matching the depth of DEEP2. In principle, we can apply $1/V_{max}$ weighting, but it is unnecessary, 
as the number of K+A galaxies is small, which only allows us to make one or two wide bins in luminosity. A $1/V_{max}$ weighting does not significantly change the effective median magnitude of K+As in each bin. In addition, the local K+A's environment distribution has no detectable luminosity dependence, as shown in \S\ref{sec:lumdep}. We therefore do not apply any magnitude-dependent weights.

To maximize the sample from SDSS while making it $B$-band limited, we empirically determined a redshift-dependent restframe $B$-band luminosity limit. Because SDSS is $r$-band limited, this $B$-band limit is set by the bluest galaxies. The limit we apply is:
\begin{equation}
M_B(z)-5\log h < {\rm max}(100 z^2-45 z-15.8, -18.8)
\end{equation}

We also select a bright subsample in SDSS with $M_B- 5 \log h < -19.8$, which is the depth of SDSS at $z=0.12$, the upper limit in redshift for our $z\sim0.1$ sample. This bright subsample for SDSS is therefore volume-limited.

For the environment analysis, we also exclude galaxies within $1 h^{-1}Mpc$ (comoving) of a survey edge or hole, along with regions on the sky with less than 80\% completeness. With these criteria, we are left with 79716 galaxies, amongst them 187 K+A galaxies.
The redshift range, effective redshift, and magnitude limits adopted for the two parent samples are summarized in Table \ref{tab:sample}. 

\begin{table*}
\begin{center}
\caption{Sample definition summary}
\begin{tabular}{c c  c c c c r r r r}
\hline\hline
Name && Redshift & Median $Z$ &  Mag Range & Median & Sample & K+A & K+A fraction & Contamination \\
     &&  Range & & ($\tilde{M}=M_B-5\log h$) & $M_B-5\log h$ & Size & Sample & & Correction\\ \hline 
SDSS-whole && 0.07-0.12 & 0.09 & $\tilde{M}<-18.8$ & -19.73& 79716 & 187 & $0.23\pm0.02\%$ & 0.05\% \\
SDSS-bright&& 0.07-0.12 & 0.09 & $\tilde{M}<-19.8$ & -20.16& 35717 & 93 & $0.26\pm0.03\%$ & 0.02\% \\
SDSS-faint && 0.07-0.12 & 0.09 & $-19.8<\tilde{M}<-18.8$ & -19.42& 43999 & 94 & $0.21\pm0.02\%$ & 0.07\%\\ \hline
DEEP2-whole&& 0.72-0.88 & 0.79 & $\tilde{M}<-19.7$ & -20.36& 2649 & 34 & $1.3\pm0.2\%$ & 0.9\% \\
DEEP2-bright&& 0.72-0.88 & 0.79 & $\tilde{M}<-20.7$ & -21.02& 783  & 17 & $2.3\pm0.5\%$ & 1.1\%\\
DEEP2-faint & & 0.72-0.88 & 0.79 & $-20.7<\tilde{M} <-19.7$ & -20.12& 1866 & 17 & $0.9\pm0.25\%$ & 0.8\%\\
\hline
\end{tabular}
\label{tab:sample}
\end{center}
\end{table*}


\subsection{K+A Galaxy Fraction and the Contamination Correction}
\label{sec:contami}
It is easily seen in Fig.~\ref{fig:deep_1p} that the K+A population in DEEP2 is not so obviously a separate population from star-forming galaxies as in SDSS. This is largely due to the much larger sample size of SDSS, but the greater measurement errors in DEEP2 spectral indices also obscure the gap. Therefore, to reliably estimate the K+A galaxy fraction, we need to do a contamination correction, i.e., estimate the expected false-positive rate and the true-negative rate for a given environment bin, then subtract or add them, respectively, to the raw K+A fraction. 

A robust contamination estimate requires reliable error estimates for both spectral indices used. The error on each index is first estimated from the inverse variance array output from the DEEP2 DEIMOS spectroscopic reduction pipeline for each object. However, there can also be systematic errors due to observational conditions, such as incomplete slit coverage, variable seeing conditions, misalignment between slit and atmospheric dispersion direction, etc., that cause pipeline errors based on measurement noise to be an underestimate of the true measurement error. To give reliable errors, we follow \cite{Balogh99} and use objects with repeated observations to scale up the pipeline errors such that repeated measurements on the same object give consistent results. The errors on \hb\ EW and $f_{\rm A}$ are scaled up by factors of 1.81 and 1.61, respectively. The details of this rescaling can be found in Yan et al. in prep. Similarly, the measurement errors on galaxies in SDSS are also scaled according to the estimates from repeated observations. 


With reliable error estimates for \hb\ EW and $f_{\rm A}$, we can predict how much contamination these errors (presumed to be Gaussian) will cause, assuming the true distribution is the same as the observed distribution. This is done by Monte-Carlo simulations. We add Gaussian noise to the observed \hb\ EW and $f_{\rm A}$ values, according to the error estimates of each point. We then calculate, for each point, the likelihood it falls into the K+A definition box. This likelihood is also the probability for each galaxy being a true K+A galaxy. The difference between the sum of the likelihoods and the raw fraction of K+A galaxies gives the contamination correction. The fact that the observed distribution differs from the true distribution (due to measurement errors) causes this contamination estimate to be imperfect. However, for distributions with small second derivatives (as here), this effect is modest and may be safely ignored. 





\subsection{Environment Measures}\label{sec:env}
The local galaxy environment has been estimated for each galaxy in the DEEP2 and SDSS samples that is at least $1 h^{-1}$ Mpc (comoving) away from a survey edge (see \citealt{Cooper06,CooperNC07} for measurements in DEEP2 and \citealt{CooperNW08} in SDSS). 
The environment indicator employed is the projected $3^{\rm rd}$-nearest neighbor surface density ($\Sigma_3$), which is derived from the projected distance to the $3^{\rm rd}$-nearest-neighbor, $D_{p,3}$ , through $\Sigma_3 = 3/(\pi D_{p,3}^2)$.  The measurement is made for objects within a velocity window of $\pm 1000{\rm km/s}$ along the line-of-sight in order to exclude foreground and background galaxies. For each galaxy, the surface density is divided by the median $\Sigma_3$ of galaxies at that redshift to correct for the redshift dependence of the sampling rate in both the DEEP2 and SDSS surveys. This converts the $\Sigma_3$ values into measures of overdensity relative to the median density (given by the notation $1+\delta_3$ here). Details of the environment measurements can be found in \cite{Cooper06, CooperNC07, CooperNW08}. 
The environment indicator employed here for the SDSS sample is different from the indicators used by \cite{BlantonEH05} and \cite{Hogg03,Hogg06}. These differences are not very significant as the rank order of environment among galaxies should be similar, and environments are measured on similar physical scales in \cite{Hogg06} and here. 
For detailed comparisons of a few different environment indicators, see \cite{Cooper05}. 


We note that this environment indicator measures different scales in different environments, which also vary with redshift. At $z\sim0.8$, simulations show that, at $\log (1+\delta_3) > 0.5$, this indicator can be roughly translated as the hosting halo mass; at $\log (1+\delta_3) < 0.5$, this indicator measures the inter-halo environment 
Typically it measures local density on a 2Mpc scale. 
Although SDSS has a higher sampling density than DEEP2, our environment indicator measures a slightly larger scale in SDSS than in DEEP2: the median $D_{p,3}$ is $\sim1.3h^{-1} Mpc$ (comoving) in the SDSS sample and $\sim1.05h^{-1}Mpc$ (comoving) in the DEEP2 sample. 
It is important to remember that this environment indicator is normalized relative to the median overdensity of all galaxies at each redshift, therefore is a relative measure of environment. This is important when we consider environment evolution with redshift.


\section{Color-Magnitude Distribution of Post-quenching Galaxies}

\begin{figure*}
\begin{center}
\includegraphics[angle=90,totalheight=0.35\textheight,viewport=0 0 420 770,clip]{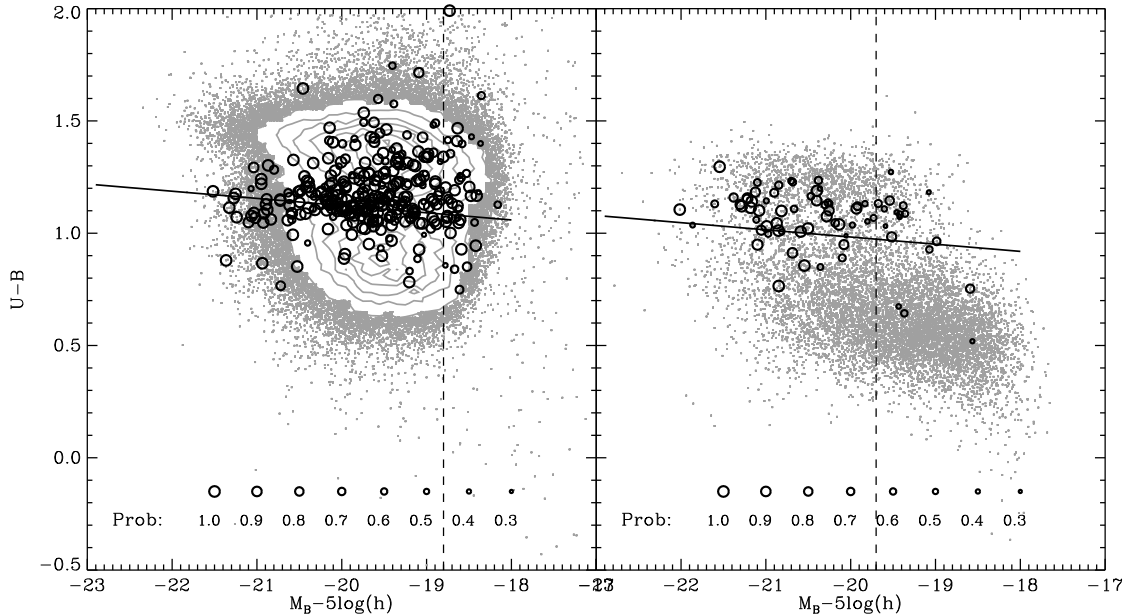}
\caption{This plot shows the colour-magnitude distributions of SDSS (left panel) and DEEP2 (right panel) K+A galaxies. The background gray contours or points represent all galaxies within the same redshift ranges with a successful redshift measurement. The gray points represent all galaxies within the same redshift range in DEEP2 with a successful redshift measurement. K+A galaxies are highlighted as circles with sizes proportional to the degree of confidence of their identification. In each panel, the vertical dashed line indicates the absolute magnitude limit adopted in each sample in all following analyses; the tilted solid line indicates the red/blue demarcation in each sample, which is 0.14 magnitude redder at the SDSS redshifts ($z\sim0.1$) than at the DEEP2 redshifts ($z\sim0.8$). For the DEEP2 sample at $M_B-5\log h$ fainter than -20.4, we apply an additional cut, S/N per pixel $ > 1$, before making the K+A selection; these fainter galaxies are plotted here, but not included in our K+A sample in the analyses below.}
\label{fig:cmd}
\end{center}
\end{figure*}

Figure~\ref{fig:cmd} shows the colour-magnitude distributions of all K+A galaxies in SDSS and in DEEP2 selected by the methods described above in \S\ref{sec:select} (circles), while the distribution of the parent sample is plotted in grey countours for SDSS and in grey points for DEEP2. 

As seen in numerous studies \citep[e.g.][]{BellWM04,Willmer06}, the restframe colour distribution of all galaxies is bi-modal at both $z\sim0.1$ and $z\sim0.8$, with a tight red galaxy sequence and a diffuse 'cloud' of blue galaxies. For DEEP2, we use the same magnitude-dependent colour cut as in \cite{Willmer06} to separate red and blue galaxies, as indicated by the tilted solid line.
\begin{equation}
U-B = -0.032(M_B-5\log h)+0.343
\end{equation} 
One might notice that the colour-magnitude distribution of SDSS galaxies is shifted redward relative to that of DEEP2. This is due to the aging of the stellar population and the decrease in the star formation rate from $z\sim0.8$ to $z\sim0.1$ \citep{Blanton06}. The red sequence is also much more densely populated at lower $z$. This echoes the observed growth of this population from luminosity function studies \citep{Faber07,Brown07}. Since galaxies are redder at low $z$, we adopt a redder colour cut for SDSS, which is 0.14 mag redder than that for DEEP2, the same as that adopted by \cite{CooperNW08}. This cut separates the red and blue galaxies at low $z$ reasonably well.

The K+A galaxies are highlighted as circles on these plots. The size of each circle indicates the corresponding galaxy's probability of being a true K+A (see \S\ref{sec:contami}). The magnitude limit for each sample is indicated by the vertical dashed line in each panel.




Despite the shifts in colour of the red sequence and the blue cloud from DEEP2 to SDSS, the median $U-B$ colour of the K+A galaxies remains largely unchanged. This is expected as they are selected in the same manner, through decomposition of the stellar spectra, thus having the same light-weighted mean stellar age. This further demonstrates the equivalence between the two K+A samples selected from the two surveys. In DEEP2, K+A galaxies mostly occupy the red side of the ``green valley'' and the blue side of the red sequence. In SDSS, they are mainly in the ``green valley'', bluer than most of the red sequence galaxies. This difference in their positions relative to the red sequence is entirely due to the aging of stellar populations on the red sequence from $z\sim0.8$ to $z\sim0.1$. 

\section{Environments of Post-quenching Galaxies}

In this section, we investigate the environment distribution of K+A galaxies. 
We face several difficulties. First, the environment measurements have large errors, predominantly due
to the sparse sampling of the matter distribution provided by a galaxy survey.
Environment differences between samples
can only be detected when the sample sizes are large. Secondly, our K+A sample size, 
although being the largest available at this redshift, is small for statistical purposes. Thirdly,
all statistical presentations and tests have limitations, especially given 
noisy measurements and small sample sizes. To interpret the results, we
have to keep these facts in mind and combine multiple methods to reach a conclusion.
Therefore, we investigate the environment distribution of K+A galaxies in several
ways.

\subsection{SDSS}\label{sec:sloan_env}



Before we look at the higher-$z$ sample from DEEP2, we first investigate the environments of K+As in the SDSS sample, which gives a larger K+A sample and better statistics.

Our first approach is to compare the K+A fractions among galaxy subsamples binned by their local overdensity. This is similar to the approach in most previous studies, which compared the fractions of K+As between two classes of environment: clusters vs. field. 

\begin{figure}
\begin{center}
\includegraphics[totalheight=0.35\textheight]{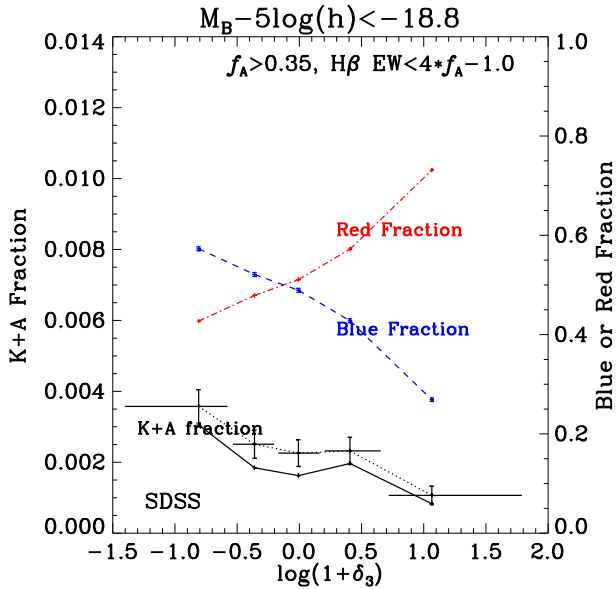}
\caption[Fraction of K+A galaxies as a function of environment in SDSS]{The fractions of K+A galaxies, blue galaxies, and red galaxies as a function of environment for the SDSS sample.
The raw K+A fraction is shown as the dotted line with error bars. The K+A fraction after applying the contamination corrections described in \S\ref{sec:contami} is shown as the solid line.
The horizontal bars indicate the 90-percentile ranges of environment for each bin. The blue dashed line and the red dash-dotted line, which are plotted with respect to the vertical axis on the right, show the blue galaxy fraction and the red galaxy fraction, respectively, in the sample as a function of environment.
The separation between blue and red galaxies is defined here using the cut (in $U-B$) applied by \cite{CooperNW08}, which is 0.14 mag redder than the DEEP2 cut to account for the colour evolution of galaxies. K+A galaxies in SDSS show a similar downward trend as blue galaxies. However, as discussed later, the K+A fraction trend in this plot depends sensitively on the sample used for the denominator, making it difficult to compare with other samples. The significance is also overestimated due correlations between the bins.
}
\label{fig:frac_envr_sloan}
\end{center}
\end{figure}

In Fig.~\ref{fig:frac_envr_sloan}, we present the K+A fractions relative to all galaxies as a function of environment in the SDSS sample. The environment bins are chosen so that each one contains an equal number of galaxies (1/5 of all galaxies in this case). 

\begin{figure}
\begin{center}
\includegraphics[totalheight=0.35\textheight]{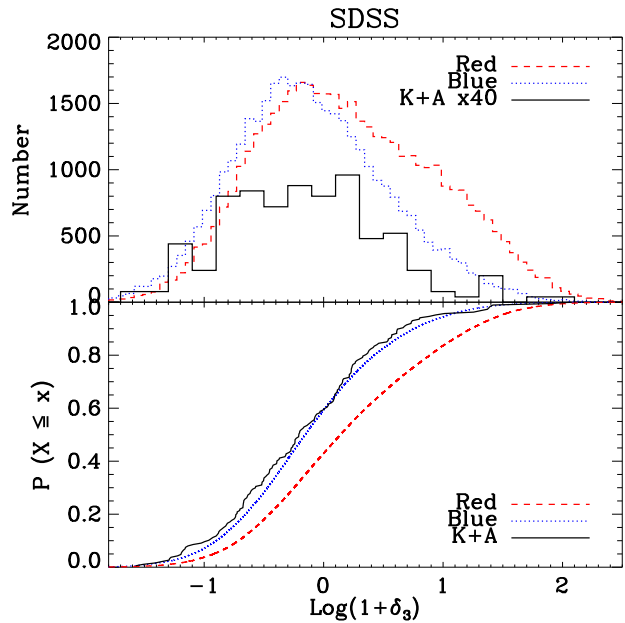}
\caption{Shown in the upper panel are the local density distributions of K+A galaxies (black, solid line), blue galaxies (blue, dotted), and red galaxies (red, dashed) in the SDSS sample. The histograms for the K+A galaxies are magnified for ease of comparison. The lower panel shows the corresponding cumulative distribution for each population. 
In SDSS, the K+A galaxies show a very similar environment distribution as blue galaxies, and is significantly different from red galaxies. 
}
\label{fig:delta3_hist_sloan}
\end{center}
\end{figure}

Since the measurement of a fraction has a binomial distribution, the error bars are computed according to the variance of a binomial distribution. For small fractions, as for K+A galaxies, this is very close to Poisson error. These errors are underestimated because all the bins are correlated with each other due to environment uncertainties. 
The errors on the contamination correction for K+A samples is assumed to be small, since it is based on empirical error estimates on spectral measurements with repeated observations. 

As shown in Fig.~\ref{fig:frac_envr_sloan}, the blue galaxy fraction decreases with increasing local density, while red galaxy fraction increases with local density, as expected from the color-density relation. The K+A fraction shows a similar downward trend as blue galaxies.

In Fig.~\ref{fig:delta3_hist_sloan}, we present the environment distributions for red, blue, and K+A galaxies using histograms and cumulative distributions. These histograms first show the level of spread in local density of these subsamples, which includes both intrinsic spread and measurement error. As expected, the blue galaxies have a distribution that is centered at a lower overdensity than red galaxies, and the K+A galaxies have a distribution that is similar. The same thing can be seen in the cumulative distributions: blue galaxies occupy lower densities than red galaxies, and the K+As' cumulative distribution is similar to that of blue galaxies, with only small differences at the tails of the distribution (of low statistical significance).

It is quite obvious that the difference between the environment distributions of blue galaxies and red galaxies is very significant, as is the difference between K+A's and red galaxies. However, is the small difference between K+As and blue galaxies significant? To address this requires a statistical test. A commonly-used non-parametric (i.e., independent of Gaussian assumption) test is the Kolmogorov-Smirnov test (hereafter K-S test), which measures the maximum distance between the two cumulative distributions. Under the null hypothesis that the two samples tested are drawn from the same population, a larger maximum distance is less likely to occur. 
Throughout this paper, we adopt a threshold significance level of 0.05: i.e., we rule out the null hypothesis only when the probability of obtaining a statistic as large as that measured or larger under the null hypothesis, the p-value, is less than 0.05. 
In Table \ref{tab:stat_whole}, we show the K-S test p-value results for each subsample. 

One drawback of the K-S test is that it is rather insensitive to the distribution tails. A powerful alternative, the Anderson-Darling test (hereafter A-D test, \citealt{AndersonD54, Pettitt76, SinclairS88}) is more sensitive to differences on the tails of the distribution. In this case, the test statistic is a weighted integration of the squared differences between the two cumulative distributions, with more weight placed on differences in the tails of the distributions. In addition, we also apply the Mann-Whitney U test (also called the Mann-Whitney-Wilcoxon, or Wilcoxon rank-sum test, hereafter MWW test, \citealt{MannW47}), which is more sensitive to differences than the K-S test when sample sizes are small. It uses the relative ranks of measurements in a combined, sorted list to test the hypothesis that two samples might come from the same population.  We present the results for all these tests in Table \ref{tab:stat_whole}. 
Under all of these tests, the null hypothesis that red galaxies and K+A galaxies are drawn from the same distribution is rejected with a significance level of $10^{-5}$ or better. 
The null hypothesis that the blue galaxies and K+A galaxies are drawn from the same distribution cannot be rejected at the 0.05 significance level, however. Therefore, we conclude that K+A galaxies at $z\sim0.1$ have an environment distribution that is indistinguishable from blue galaxies, given our samples.

\begin{table*}
\begin{center}
\caption{Statistical Tests of Differences Between Samples.
This table lists the statistical tests results on the differences between the two samples in each pair. The values listed are the probabilities of obtaining a test statistic as large as that measured or larger under the null hypothesis that the two samples are drawn from the same population. This probability is called the p-value for most statistical tests. We use a threshold signficance level of 0.05, i.e., we rule out the null hypothesis only when the p-value is less than 0.05.
}
\begin{tabular}{l c c c c}
\hline
Sample & subsamples & Kolmogorov-Smirnov & Anderson-Darling &  Mann-Whitney U test\\ \hline
SDSS &red vs. blue&  $<10^{-5}$ & $<10^{-5}$ & $<10^{-5}$\\
     &red vs. K+A &  $<10^{-5}$ & $<10^{-5}$ & $<10^{-5}$\\
     &blue vs. K+A& 0.419 & 0.306 & 0.246\\ \hline
DEEP2&red vs. blue&  $<10^{-5}$ & $<10^{-5}$ & $<10^{-5}$\\
     &red vs. K+A & 0.739 & 0.912 & 0.937\\
     &  (weighted)& 0.980 & 0.946 & \\
     &blue vs. K+A& 0.166 & 0.120 & 0.080\\ 
     &  (weighted)& 0.116 & 0.049 & \\
\hline
\end{tabular}
\label{tab:stat_whole}
\end{center}
\end{table*}

\subsection{Comparison of Methodologies}
All the methods presented above -- fraction vs. environment plots, density distributions, cumulative distributions, and statistical tests -- lead to the same conclusion. The K+A galaxies at $z\sim0.1$ in SDSS have an environment distribution which is distinct from that of red galaxies, but similar to that of blue galaxies. 

The first method, the fraction vs. environment plot, is the easiest to read but the hardest to interpret and compare between different samples. First, it is very easy to overestimate the significance of the trend presented in such a plot, because the fractions in all the bins are correlated. For example, in Fig.~\ref{fig:frac_envr_sloan}, it looks quite convincing that the K+A fraction drops consistently, except for the second to last bin. In fact, 
the middle three bins are basically consistent with being flat. Their relative heights change significantly if we shift the bins slightly.
The most significant variation in the K+A fractions is the difference between the leftmost bin and the rightmost bin: the latter is significantly lower. 

Secondly, the slope of the trend sensitively depends on the relative mix of blue and red galaxies in the sample we are taking a fraction of. 
If we increase the overall fraction of red galaxies without changing the K+A population -- for instance, by switching to an $I$-band selected sample -- we will include more galaxies in overdense environments. Under the same equal-number binning, the bin boundaries will shift to higher densities. Even if the K+A population included does not change, the slope of the K+A fraction will still become steeper, i.e., decrease faster towards denser environments. The lesson is that the slope of the fraction vs. environment trend is sensitive to the parent sample, which means we cannot blindly compare the K+A fraction slopes between different samples, such as ones defined using different limiting band, using different luminosity cuts, or at different redshifts. We will see an example of this when we investigate luminosity-dependence (\S\ref{sec:lumdep}) below. 

The second presentation method used is the density distribution histogram. The histograms provide much more detail than the fraction plot, but for small samples like K+A galaxies, the number of bins has to be relatively small, thus limiting the power of the data. Cumulative distributions give a much more sensitive presentation: it does not require binning, and thus can take full advantage of the data while making the presentation as continuous as possible. We therefore consider this presentation the most accurate and informative.

In fact, many of the non-parametric statistical tests are based on comparisons between empirical cumulative distribution functions (CDF). They provide quantitative tests of the significance of the differences between different samples. They also have limitations, however. Different tests compare different aspects of the data, thus they do not necessarily give the same answer. The test giving the smallest p-value is not necessarily the most powerful. This will be clear in the following analysis for DEEP2 K+As.

Comparing all these methods, we consider the cumulative distribution the most accurate and informative way of presentation, and we base our judgements on the statistical tests for our conclusions.

Now we move on to investigate the environments of K+A galaxies in the DEEP2 sample, which is a smaller sample and hence more noisy.

\subsection{DEEP2}
\begin{figure}
\begin{center}
\includegraphics[totalheight=0.35\textheight]{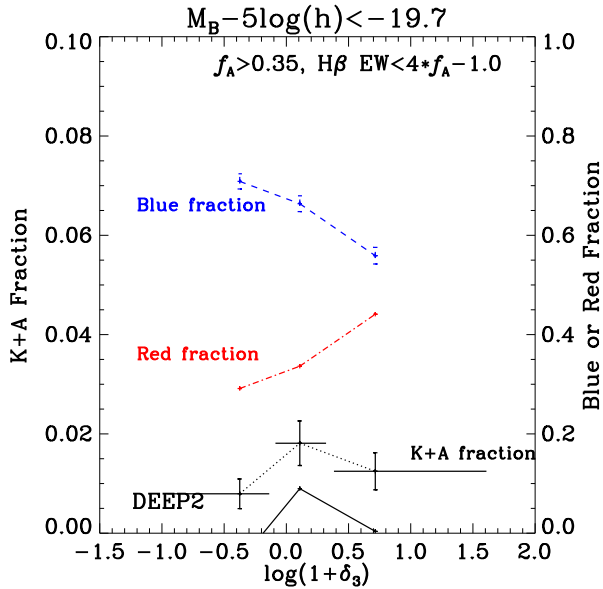}
\caption[Fraction]{The fractions of K+A galaxies, blue galaxies and red galaxies as a function of environment for the DEEP2 sample. All symbols follow the same conventions as in Fig.~\ref{fig:frac_envr_sloan}. The separation between blue and red galaxies is defined as in \cite{Willmer06}. The error bars for the blue and the red fractions are the same since they sum to 1. K+A galaxies appear to have a slight preference for the central environment bin. However, the significance of this trend is plagued by small number statistics and is less robust.}
\label{fig:frac_envr}
\end{center}
\end{figure}

As in \S\ref{sec:sloan_env}, we present the K+A environment distribution in DEEP2 to blue and red galaxies with three different methods. Figure \ref{fig:frac_envr} shows the K+A fraction as a function of environment, along with that for blue galaxies and red galaxies. The blue fraction decreases toward denser environments and the red fraction increases, as expected. The K+A fraction shows a slight preference for the central environment bin (although the contamination correction is large for the DEEP2 sample, the trend does not change with or without the correction).
This indicates its environment distribution is in between that of the blue and red samples. However, due to the correlation among the three bins and the small sample involved, the K+A trend presented by this plot is not very robust. We turn to the density distributions and cumulative distributions for deeper analysis.

\begin{figure}
\begin{center}
\includegraphics[totalheight=0.35\textheight]{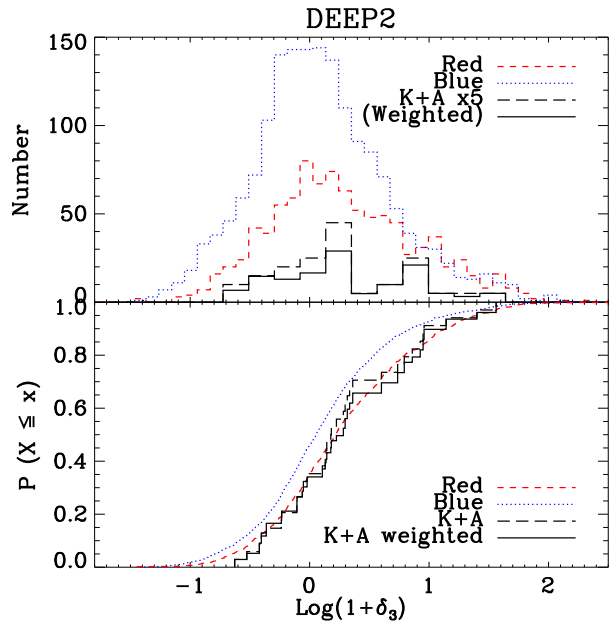}
\caption{Shown in the upper panel are the local density distributions of K+A galaxies (black, solid line), blue galaxies (blue, dotted), and red galaxies (red, dashed) in the DEEP2 sample. The histogram for the K+A galaxies is magnified for ease of comparison. The lower panel shows the corresponding cumulative distribution for each population. In DEEP2, the K+A galaxies' environment distribution is in between those of red and blue galaxies, perhaps more similar to blue galaxies than to red galaxies. 
}
\label{fig:delta3_hist}
\end{center}
\end{figure}

Figure \ref{fig:delta3_hist} shows the density distributions and cumulative distributions for red, blue, and K+A galaxies in the sample. For K+A galaxies, we make two sets of density distribution and cumulative distribution: one set is based on pure number counts, the other set is weighted by the probability of each K+A being a true K+A galaxy. We do this because the spectral measurements in DEEP2 have larger errors than those in SDSS and the sample size is small. Thus, giving more secure K+A galaxies higher weight can make the result more robust. 

As seen in this figure, K+A galaxies show a density distribution skewed towards overdense regions, more similar to red galaxies than blue. The cumulative distributions basically tracks that for red galaxies, especially when weighted by K+A probability. The statistical tests also show that the environment distribution of K+A galaxies is more similar to red galaxies than to blue galaxies, as seen in Table~\ref{tab:stat_whole}. K+A galaxies are consistent with being drawn from the same population as red galaxies in all tests, regardless of weighting. In the case for blue galaxies, the probabilities are approaching the threshold for rejecting the null hypothesis. However, the differences do not reach our significance threshold ($p<0.05$) except for the weighted A-D test (p=0.049). Thus, we cannot safely conclude that K+A galaxies at $z\sim0.8$ have different environment distribution as blue galaxies. We can only conclude that they are possibly more similar to red galaxies than to blue galaxies. The difference become more significant when we weight more secure K+As more.

Small sample size and large intrinsic error in the environment measures both affect the significance of the results. To improve the significance, larger samples and/or more accurate environment measures are required.

\subsection{K+As in Groups vs. Fields in DEEP2}

Another way to characterize the local environment of galaxies is to classify galaxies in one of two classes: cluster/group members and isolated field galaxies. Here we use the DEEP2 group catalog produced by \cite{Gerke05} to investigate the K+A fractions in the DEEP2 groups and in the field. We note here that few of our galaxies are in large clusters, because the volume of DEEP2 is not large enough and there are fewer clusters at $z\sim1$ than $z\sim0$.

The group catalog used here is an expanded version of the one presented in \cite{Gerke05}, using the complete DEEP2 datasets rather than the first 25\% data. The groups are identified using the Voronoi-Delaunay Method \citep{Marinoni02}. For details of the method, along with assessments of the completeness and purity of the resulting group catalog, see \citet{Gerke05}. We calculate the K+A fraction in galaxy groups with estimated velocity dispersion of $\sigma > 200$km/s, which corresponds to a minimum dark matter halo mass of $\sim6\times10^{12}h^{-1} M_\odot$ \citep{CoilGN06}. Out of 671 galaxies with $M_B-5\log h$ brighter than $-19.7$ in 329 groups within the redshift range considered, the raw K+A fraction is $1.0\pm0.4\%$. Among 1613 field galaxies outside groups, the raw K+A fraction is $1.5\pm0.3\%$. Applying contamination corrections, the true K+A fractions in groups and in the field are $0.1\pm0.4\%$ and $0.7\pm0.3\%$, respectively. 
We also list the K+A fractions in groups by their richness in Table~\ref{tab:group}. We reach same conclusions regardless of the variations in the group sample: the fraction of K+As among field galaxies is slightly higher than that among group galaxies, at 1-$\sigma$ level. The significance of the result is limited by the small number of K+As. The contamination corrections are likely overestimates, which makes the net fraction become exceedingly small or even negative. 


\begin{table*}
\caption{K+A Fractions in Groups vs. Fields in DEEP2}
\begin{center}
\begin{tabular}{l c c c c}
\hline
Description & Raw K+A Fraction & Corrected K+A Fraction & No. of Galaxies & No. of Groups \\ \hline
Groups w/ ${\rm V}_{disp}>200{\rm km/s}$ & $1.04\pm0.39\%$ & $0.13\%$ & 671 &329\\
Groups w/ 4 or more members & $0.95\pm0.36\%$ & -0.13\% & 737 & 313 \\
Groups w/ 2 or 3 members & $0.87\pm0.35\%$ & -0.20\% & 690 & 486 \\
Field galaxies & $1.49\pm0.30\%$ & $0.70\%$ & 1613 & \\
\hline
\end{tabular}
\label{tab:group}
\end{center}
\end{table*}


\subsection{Comparison with Previous Studies}

Numerous authors have studied the environment of K+A galaxies in the past. Here we review them and compare their results with ours. As we mentioned before, the environment preference of K+A galaxixes depends on the methodolgy employed, the parent sample used, and the selection critieria of K+As. In addition, different studies often adopted different environment indicators. All these make it difficult to do a fair comparison. Nonetheless, we review the results from a few previous studies here.

\subsubsection{Low Redshift}
At low redshift, nearly all studies reached the same conclusion as ours: K+As at low-redshift have a similar environment distribution as blue galaxies, which are predominantly found in the field. 

\cite{Zabludoff96} first did this study at low redshift in the Las Campanas Redshift Survey. With a K+A selection based on strong Balmer absorption lines and lack of \oii\ emission, they found that at least 75\% of K+A galaxies are found in the field, not in or near clusters.
\cite{BlakePC04} used the 2dF Galaxy Redshift Survey and adopted a similar selection criteria to study the environment of K+A galaxies. They found the low-redshift K+A galaxies are located predominantly in the field. These two studies both selected K+A based on the lack of \oii\ emission, which would make the sample incomplete for K+As hosting a low-luminosity AGN. 

\cite{Quintero04} studied the K+A environment in SDSS and used a selection method very similar to ours.\footnote{In place of our cut in \hb\ EW and $f_A$ space, \cite{Hogg06} used \hal\ EW and A/K ratio to select K+As.}  They found that the mean overdensity of K+As on the 1 and 8 Mpc/h scales are similar to the mean of spiral galaxies, most of which live in the field, and is lower than the mean of bulge-dominated galaxies, many of which occupy high density environment. With the same K+A sample, \cite{Hogg06} employed three different environment indicators: the number density in 8 Mpc/h comoving spheres, the transverse distance to the nearest Virgo-like cluster, and the transverse distance to the nearest luminous neighbour. They found that K+A galaxies at low $z$ have a very similar environment distribution to star-forming galaxies, with only small, statistically insignificant differences. We confirm these results at low z. 

\cite{Goto05} selected K+A galaxies in SDSS with strong \hd\ absorption but without either \hal\ or \oii. The additional requirement of not having \oii\ will make the sample more restrictive than what we define here. With this seleciton, they found the environment of K+As at 0.5, 1.5 and 8Mpc scales are consistent with that of field galaxies, and inconsistent with that of cluster galaxies. With basically the same sample, \cite{Balogh05} used the $\Sigma_5$, derived from the projected distance to the fifth-nearest neighbour brighter than $M_r = -20$, to characterize the environment. They concluded that K+A galaxies reside in environment typical of normal SDSS galaxies and that are inconsistent with overdense regions like clusters.


\subsubsection{High Redshift ($0.3<z<0.9$)}
At redshift greater than 0.3, the environment of K+A galaxies is an unsettled question. But all the high-$z$ studies done in the past used \oii\ as the star formation indicator to select K+A galaxies. As mentioned in \S\ref{sec:select} and discussed in \cite{Yan06}, \oii-based selection methods suffer from incompleteness caused by AGN line emission and contamination from dusty star-forming galaxies. This can be particularly bad in clusters since \oii\ can be strong in LINERs most of which reside in early-type galaxies \citep{HoFS97V,Yan06}. 
Besides this selection difference, to compare our results with previous work also requires matching the all-galaxy sample. This is critical and was often ignored. All these difficulties prevent us from doing a fair quantitative comparison. Nonetheless, we list the results from a few representative works here.

\cite{Dressler99}, \cite{Poggianti99} studied the K+As in 10 rich galaxy clusters in the range $0.37<z<0.56$. They found that the K+A fraction in clusters ($\sim18\%$) is significantly higher than that in the field ($\sim2\%$). \cite{TranFI03} studied K+As in 3 clusters at $z=0.33$, 0.58, and 0.83 and determined that the K+A fractions in them are between 7-13\%. \cite{TranFI04} quantified the K+A fraction in the field sample from the same survey to be $2.7\pm1.1\%$, thus they concluded that K+A fraction is significantly higher in clusters than in the field. However, \cite{Balogh99} used CNOC1 survey and found similar K+A fractions between cluster and field environments. The difference from these results are likely due to differences in cluster selection (optical selection vs. X-ray selection) and in galaxy sample selection (morphological selection vs. magnitude limited sample), which are discussed in detail in \cite{Balogh99}. 
Although we do not have any Coma-like rich cluster covered in our sample, we do find the fraction of K+As in groups is similar to, perhaps slightly smaller than, the fraction outside groups. This is roughly consistent with the result of \cite{Balogh99}. However, we cannot make finer comparisons due to selection differences and lack of large clusters in our sample. 

Recently, \cite{Poggianti09} studied K+A galaxy fractions in clusters, groups, and fields using data from the ESO Distant Cluster Survey \citep{WhiteCS05}. Using a K+A selection based on \oii\ emission and \hd\ absorption, they found that the K+A fraction is higher in clusters and a subset of groups with low fraction of \oii\ emitters, than in the field. Albeit the selection differences, the K+A fraction trend they found are actually consistent with our findings. If we count both their high-\oii\ and low-\oii\ groups, the K+A fraction among all group galaxies ($5\pm3\%$) is comparable to the fraction in the field ($6\pm3\%$) (Table 4 of \citealt{Poggianti09}). This is consistent with the comparable raw K+A fractions we find among DEEP2 group galaxies ($1.0\pm0.4\%$) and field galaxies ($1.5\pm0.3\%$). The larger fractions in \cite{Poggianti09} than ours are probably due to their more inclusive selection criteria and different all-galaxy sample definition. We are unable to compare to their K+A fractions in clusters due to the paucity of rich clusters in the DEEP2 volume. The Red-Sequence Cluster Survey (RCS), with its spectroscopic follow-up program, might provide an ideal parent sample for such studies \citep{Yee07}.

\section{Testing for Luminosity Dependence of the Environment Distribution} \label{sec:lumdep}

\begin{figure*}
\begin{center}
\includegraphics[totalheight=0.35\textheight]{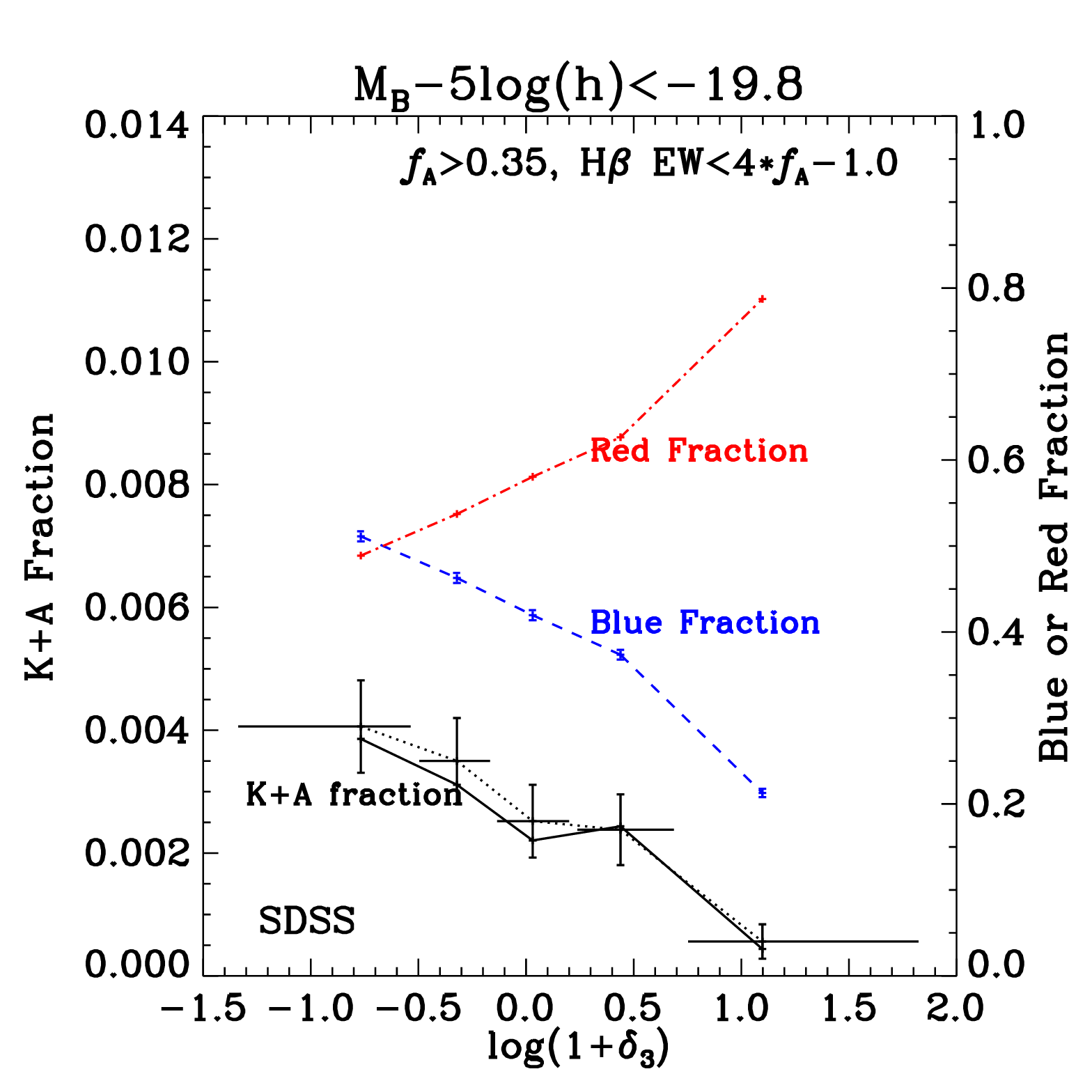}
\includegraphics[totalheight=0.35\textheight]{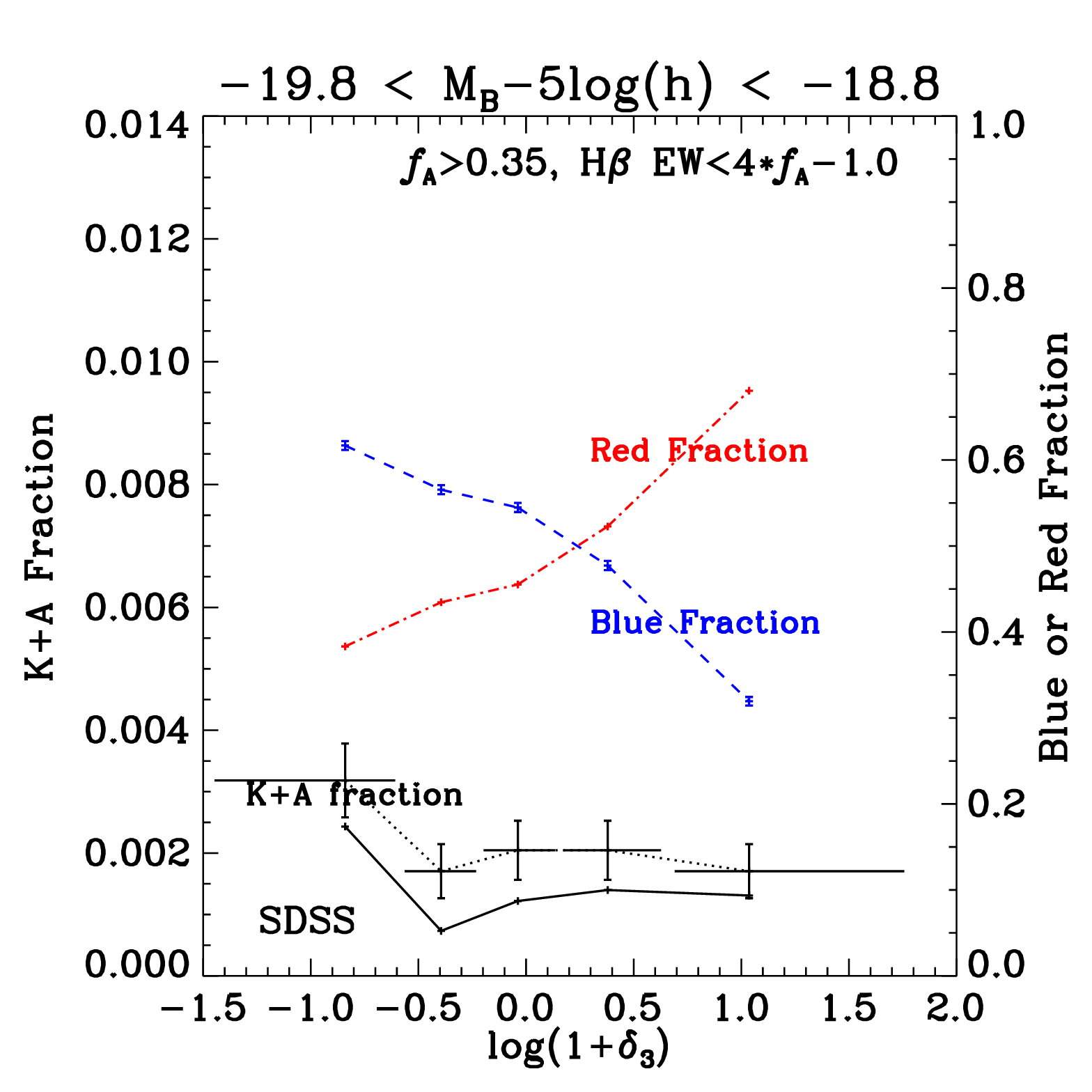}
\caption[Fraction of K+A galaxies as a function of environment in SDSS]{The fractions of K+A galaxies, blue galaxies, and red galaxies as a function of environment for the SDSS-bright sample (left) and the SDSS-faint sample (right). All symbols follow the same conventions as in Fig.~\ref{fig:frac_envr_sloan}. 
}
\label{fig:frac_envr_sloan_lum}
\end{center}
\end{figure*}

In this section, we test for luminosity dependence in the K+A galaxies' environment distribution. We also illustrate some serious drawbacks of relying on measurements of K+A fraction as a function of environment.

In Fig.~\ref{fig:frac_envr_sloan_lum}, we show the K+A fractions relative to all galaxies as a function of environment for two SDSS samples binned by luminosity: the SDSS-bright sample ($M_B < -19.8$) and the SDSS-faint sample ($-19.8 < M_B < -18.8$). Comparing the two panels, the blue fraction trend and the red fraction trend do not change greatly with luminosity, except for the overall fractions of each type. However, the K+A fraction vs. local density trends appear quite different in the two samples. For the bright sample, K+A fraction decreases with increasing local density, mimicking the environmental trend of blue galaxies. For the faint sample, K+A fraction is fairly flat across all environment bins, unlike either the red galaxies or the blue galaxies. Does this mean there is a luminosity dependence in the K+A environment distribution? The answer is no, for interesting reasons.


Contrary to one's intuition, the apparent luminosity dependence of the K+A fraction trend {\it does not} necessarily reflect a change in K+A galaxies' environment distribution with luminosity. In fact, it is largely due to the change in the environment distribution of the parent sample used to calculate fractions. Fraction measurements are dependent on the environment distribution of the all-galaxy sample used, not just the K+As. In the faint sample, there is a higher fraction of blue galaxies and a lower fraction of red galaxies than in the bright sample. The median overdensity of the sample decreases slightly; since we require that each bin has a equal number of galaxies, the bin limits shift slightly to the left. This slight shift can change the apparent trend significantly. 

\begin{figure}
\begin{center}
\includegraphics[totalheight=0.35\textheight]{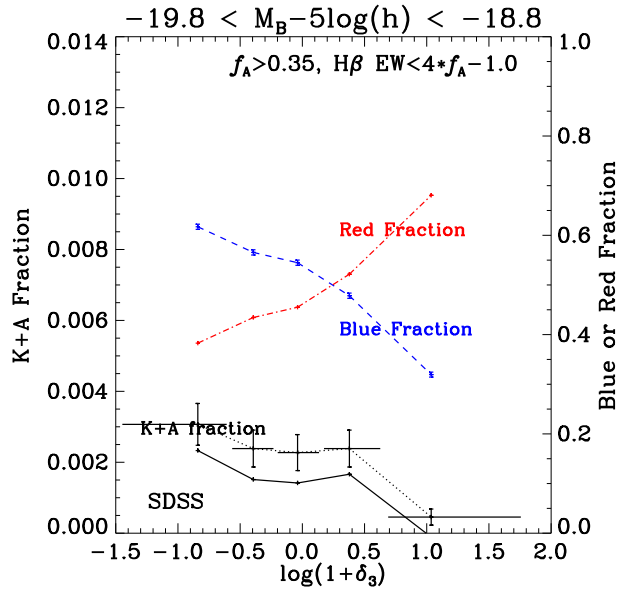}
\caption{This plot demonstrates the sensitivity of the K+A fraction slope on the parent sample. It is made with the SDSS-faint sample, but with the K+A galaxies in it replaced with those in the SDSS-bright sample. Comparing this plot with the left panel of Fig.~\ref{fig:frac_envr_sloan_lum}, which have the same K+A sample but different parent sample, we see the K+A fraction slope changed significantly.}
\label{fig:frac_envr_sloan_rep}
\end{center}
\end{figure}

We demonstrate this by replotting the panel for the faint sample, but this time replace the K+A galaxies in the faint sample by those in the bright sample. Therefore, comparing to the bright-sample plot (left panel of Fig.~\ref{fig:frac_envr_sloan_lum}), the K+A galaxies are the same, only the parent sample changed. This is shown in Fig.~\ref{fig:frac_envr_sloan_rep}. Although the K+A galaxies included are the same, the K+A fraction trend displayed is different, appearing flat except for the last bin. This demonstrates that the trend displayed in a fraction vs. environment plot is sensitive to the comparison sample. Comparing this with the original SDSS-faint sample (right panel of Fig.~\ref{fig:frac_envr_sloan_lum}), the only difference that may be significant is the K+A fraction in the last bin. However, we cannot be sure since the error bars are underestimates due to correlations amongst the bins. 

\begin{figure}
\begin{center}
\includegraphics[totalheight=0.35\textheight]{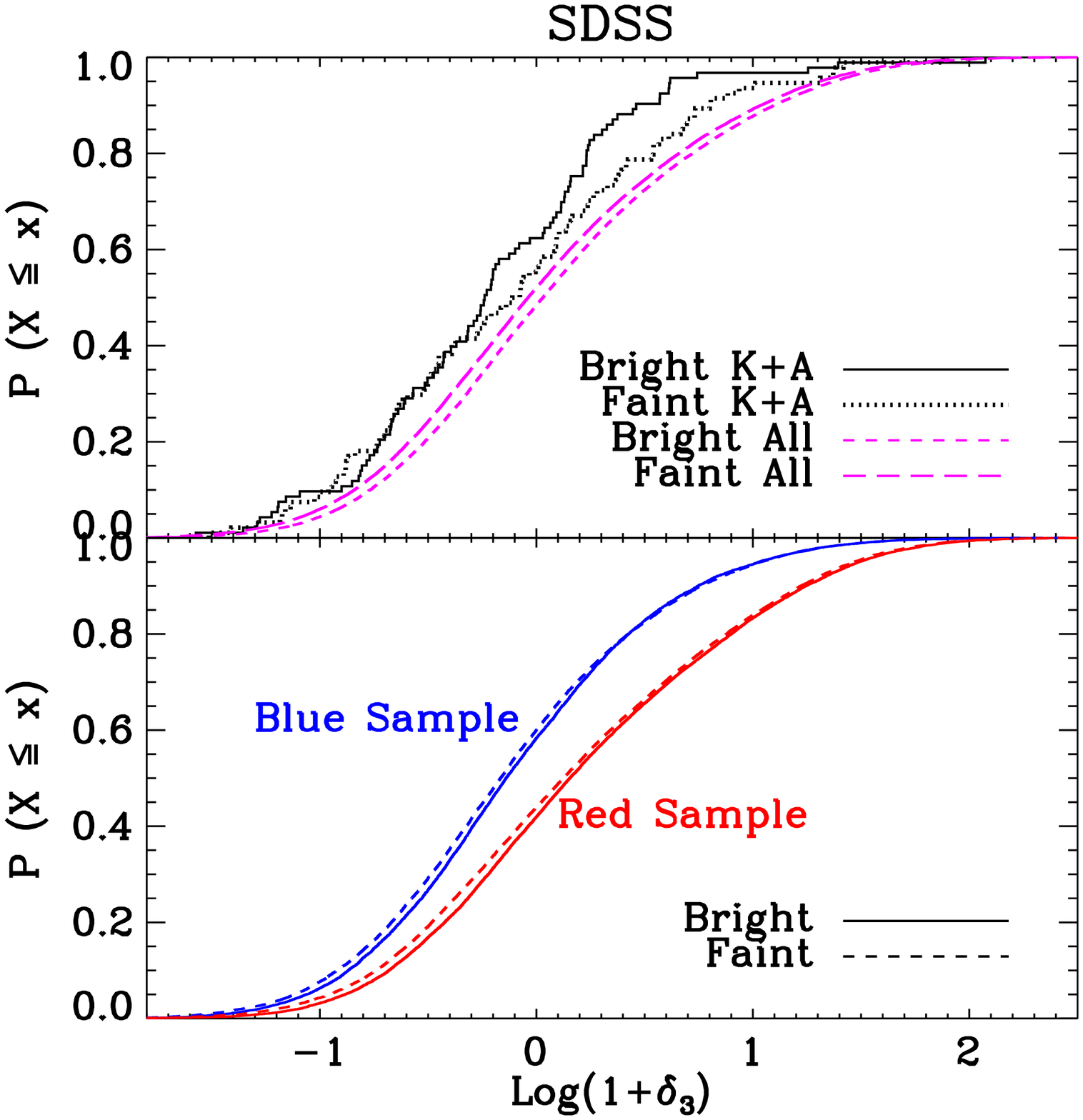}
\caption{Shown in the upper panel are the cumulative density distributions of K+A galaxies and all galaxies in the SDSS-bright and SDSS-faint samples. Due to the small samples of K+As when split in the manner, the difference between the two K+A curves are not statistically significant (see Table.~\ref{tab:stat_lum}): they are consistent with being drawn from the same population. The difference between the two all-galaxy curves, albeit small, is very significant. This is the major cause for the slope difference in the K+A fraction trend between the bright and faint sample. 
The lower panel shows the cumulative overdensity distributions for the blue galaxies (colored blue, on the left) and red galaxies (colored red, on the right) in the SDSS-bright (solid) and SDSS-faint (dashed) samples. This nicely illustrate that the environment distribution depends on both galaxy color and luminosity, but the dependence on the former dominates. 
}
\label{fig:delta3_hist_sloan_lum}
\end{center}
\end{figure}

In Fig.~\ref{fig:delta3_hist_sloan_lum}, we show the cumulative distributions of environments of the K+As, blue galaxies, and red galaxies in the SDSS-bright and SDSS-faint samples. The two K+A CDFs track each other closely in the low-density regime. There is an excess in the bright K+A sample at $\log (1+\delta_3)=-0.2$; above that, the two curves still follow each other in slope, until the faint K+A sample catches up around $\log (1+\delta_3)=1.35$. The two all-galaxy CDFs show very small but statistically significant differences. It is precisely this small difference between the two parent samples that shifts the binning and makes the slopes of the K+A fraction trends appear different. 

The problem, though related to the way in which samples are binned, is not solvable by binning in a different manner. If, instead of using equal-number binning, we fix the binning between the bright sample and the faint sample, then the K+A fraction trends can be compared with each other. However, the proportion of the parent sample in each bin would differ between samples. Thus, any difference in the resulting K+A fraction trends may not reflect the difference between two K+A samples, but the difference between the two parent samples. 

Is there any real luminosity-dependence in the K+A environment distribution? Due to the large error in the environment measurements, which makes the fraction plots useless, we have to rely on statistical tests to answer this question. Table.~\ref{tab:stat_lum} shows the p-values from the same three statistical tests used above. The environment distributions of bright K+As and faint K+As are consistent with being drawn from the same population.

\begin{table*}
\begin{center}
\caption{Statistical Tests of Luminosity Dependence in Environment Distributions}
\begin{tabular}{l c c c c}
\hline
Sample & bright vs. faint& Kolmogorov-Smirnov & Anderson-Darling &  Mann-Whitney U test\\ \hline
SDSS & All &  $<10^{-5}$ & $<10^{-5}$ & $<10^{-5}$\\
     & red &  $<10^{-5}$ & $<10^{-5}$ & $<10^{-5}$\\
     & blue&  $4\times10^{-5}$ & $<10^{-5}$ & $4\times10^{-5}$\\
     & K+A & 0.321 & 0.236 & 0.333 \\ \hline
DEEP2& All &  $6\time10^{-5}$ & $<10^{-5}$ & $<10^{-5}$\\
     & red &  $0.010$ & $0.0057$ & $0.0053$\\
     & blue&  $0.044$ & $0.0057$ & $0.0057$\\
     & K+A & 0.387 & 0.213 & 0.823\\ 
\hline
\end{tabular}
\label{tab:stat_lum}
\end{center}
\end{table*}

\begin{figure*}
\begin{center}
\includegraphics[totalheight=0.35\textheight]{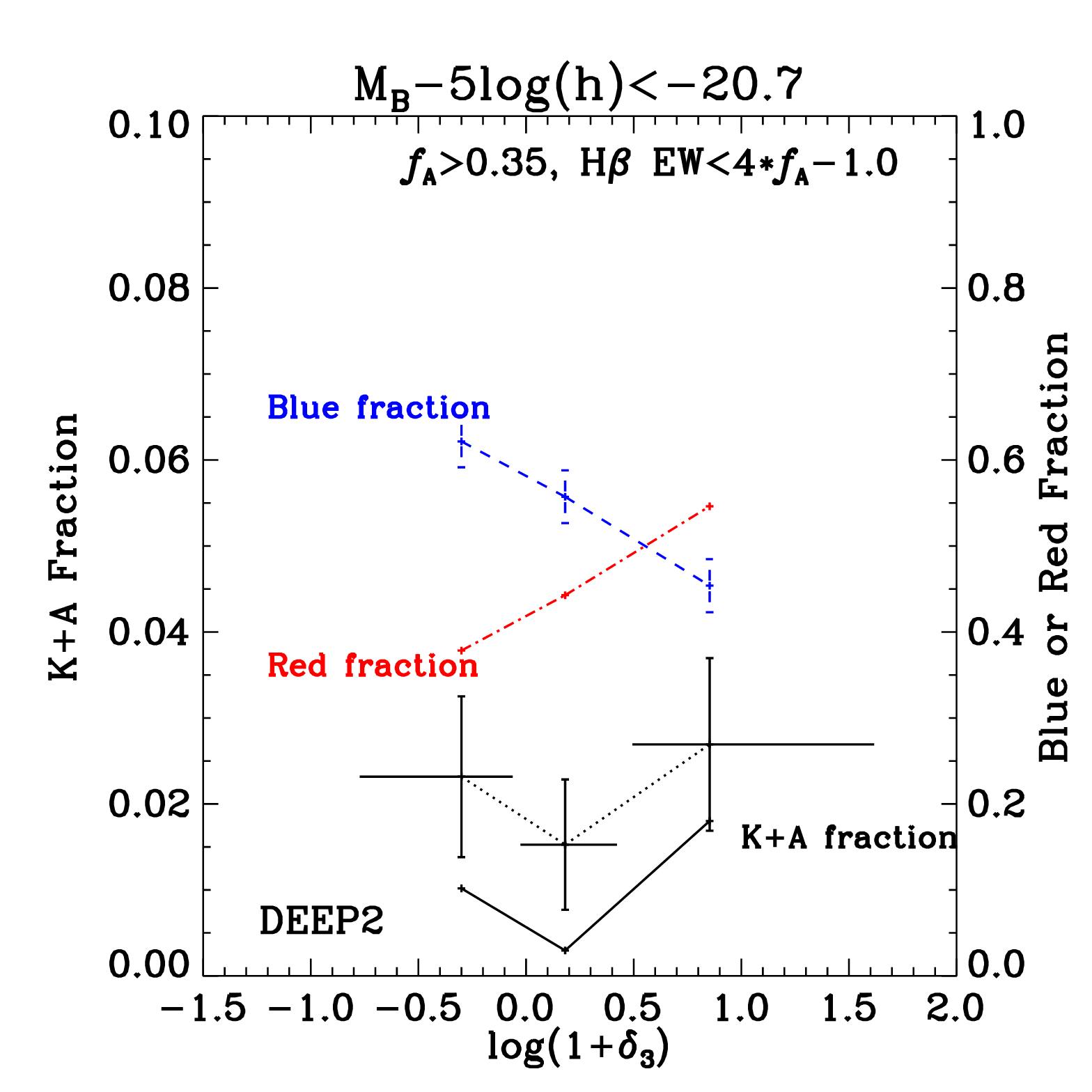}
\includegraphics[totalheight=0.35\textheight]{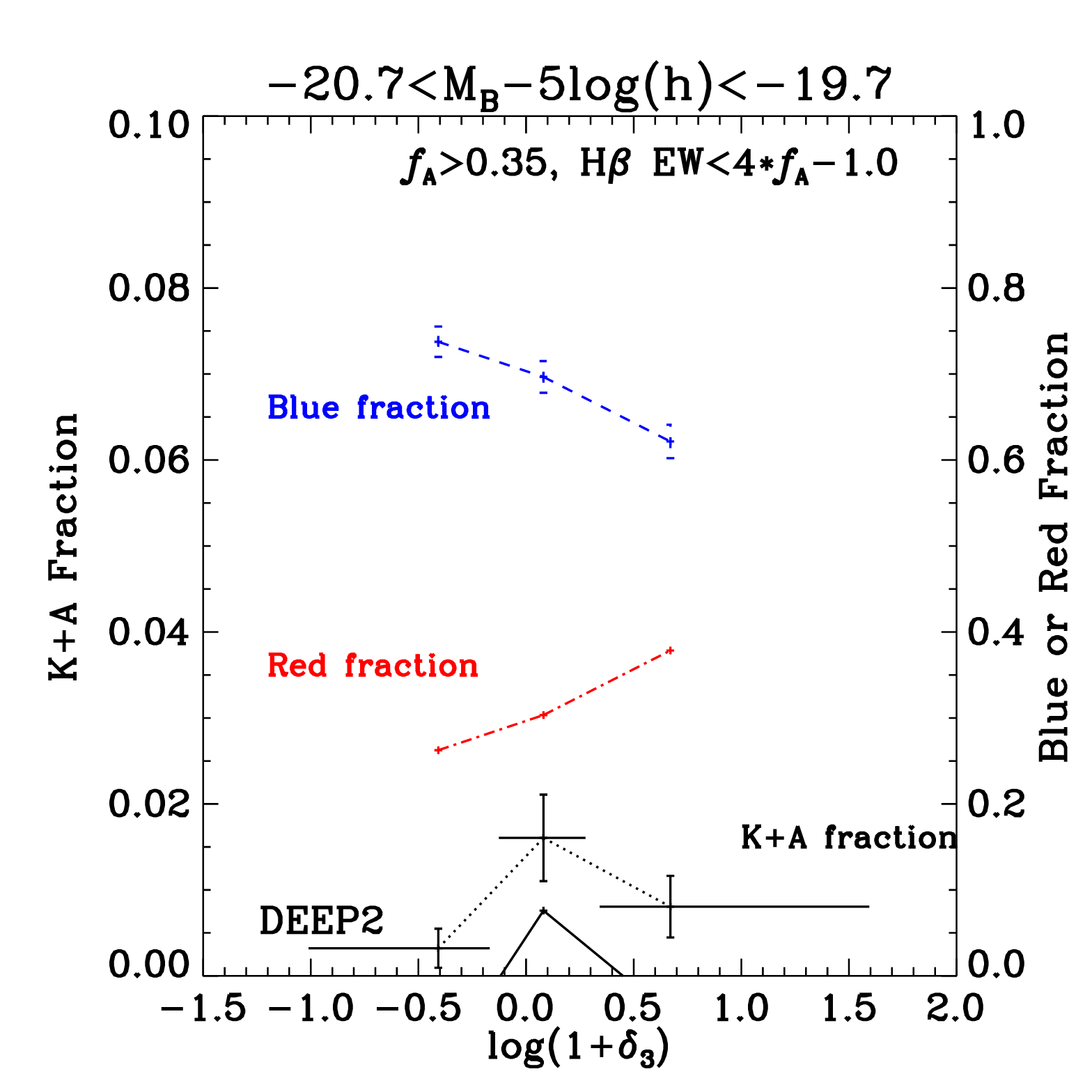}
\caption{The fractions of K+A galaxies, blue galaxies and red galaxies as a function of environment for the DEEP2-bright sample (left) and the DEEP2-faint sample (right). 
}
\label{fig:frac_envr_lum}
\end{center}
\end{figure*}

\begin{figure}
\begin{center}
\includegraphics[totalheight=0.35\textheight]{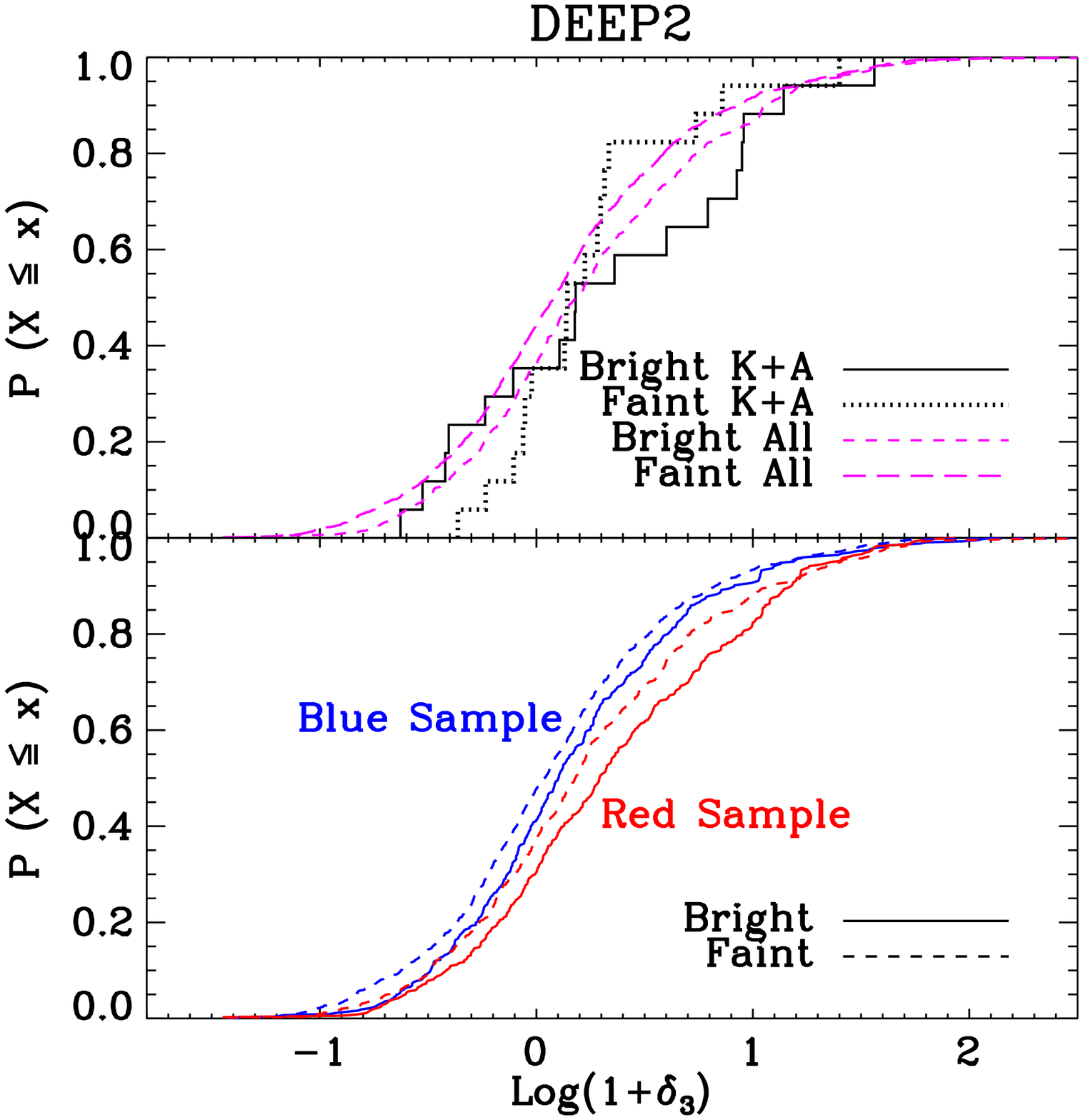}
\caption{Shown in the upper panel are the cumulative density distributions of K+A galaxies and all galaxies in the DEEP2-bright and DEEP2-faint samples. Due to the small samples of K+As when split by luminosity, the difference between the two K+A curves are not statistically significant (see Table.~\ref{tab:stat_lum}): they are consistent with being drawn from the same population. The difference between the two all-galaxy curves is highly significant. This is a major cause for the difference in the K+A fraction trend vs. environment between the bright and faint samples. The lower panel shows the cumulative distributions for the blue galaxies (colored blue, on the left) and red galaxies (colored red, on the right) in the DEEP2-bright (solid) and DEEP2-faint (dashed) samples. This nicely illustrates that the environment distribution depends on both galaxy color and luminosity, but the dependence on the former dominates. 
}
\label{fig:delta3_hist_lum}
\end{center}
\end{figure}

We repeat this exercise for the DEEP2 sample. We show the fraction plots in Fig.~\ref{fig:frac_envr_lum} and the cumulative distribution plots in Fig.~\ref{fig:delta3_hist_lum}. Again, the difference in K+A fraction trends appear at first to be very significant, but actually is not. The two K+A samples are consistent with being drawn from the same population based on all statistical tests used (see Table~\ref{tab:stat_lum}). However, for the two all-galaxy samples, there is significant luminosity dependence in their environment distributions. This luminosity dependence of the all-galaxy sample is much stronger than that in the SDSS, consistent with the results shown in \cite{Cooper06}.

In summary, with the current samples, we {\it do not} find a significant luminosity dependence in the environment distribution of K+A galaxies at either $z\sim0.1$ or $z\sim0.8$. 

\bigskip

The cautionary lesson we learned here also applies to other works looking at K+A fractions as a function of environment.
It is vital, when comparing results between different measurements, to ensure that the samples out of which K+A fractions are defined are closely matched.
Addtionally, more care must be taken to assess the significance of these results if the environment indicator used has large uncertainties. These issues may largely explain the differences in past K+A environment results in the literature.

\section{Discussion}\label{sec:discuss}
The generic quenching picture we assume here is this: a blue, star-forming galaxy undergoes an undetermined mechanism or mechanisms which may or may not depend on environment; the star formation in the galaxy is quenched and the galaxy appears as a K+A galaxy; it becomes red in $\sim1$ Gyr and evolves passively afterwards. 

\subsection{Redshift Evolution of Environments}
We have seen above that the environment distribution of post-quenching galaxies at low-$z$ is similar to blue galaxies and dramatically different from red galaxies, while the equivalent population at high-$z$ has an environment distribution more similar to red galaxies than to blue galaxies. In this section, we try to understand this. 

First, we want to remind our readers that the environment measure we adopted is a relative environment measure. It is computed relative to the median environment of all galaxies at each redshift. It is thus different from an absolute environment measure, such as the dark matter halo mass, or the galaxy number density inside a comoving volume centered on the galaxy in question. In other words, the relative environment is similar to ranking in environment, with galaxies in relatively overdense environments having higher ranks. This is important for understanding the difference in environment distribution between high and low redshift. Identical relative environments, i.e., same ranks, at different redshifts do not correspond to the same absolute environments. 

For the discussion of quenching mechanisms, we shall primarily focus on the absolute environment. However, there is one unique feature of relative environment. Galaxies are less mobile in relative-environment space than in absolute-environment space. For example, a galaxy may move into a higher density environment (absolute), but its environment rank may not change significantly, not more than the uncertainty associated with the measurement. This is because, statistically, all those galaxies with the same rank have an equal probability to have their local densities increased over the same time period. Therefore, supposing we can measure the environment to arbitrarily high accuracy, then we can assume that the environment ranks are fixed statistically for all galaxies from $z\sim1$ to $z\sim0$. In reality, we cannot measure environment to arbitrary accuracy, and due to the stochasticities involved in this measurement, there will be shuffling of ranks among galaxies with time. In addition, once the local galaxy density becomes so high that we can no longer distinguish differences from redshift-space data, for instance, when a group merges into a cluster, the ranks hit a ceiling and become meaningless. However, on a scale coarser than the measurement uncertainty, the average ranks for galaxy populations from different environments will be fixed. 

Because environment rank cannot change significantly over time, it can be very useful in identifying progenitor-descendant relationships. For two populations to have a progenitor-descendant relationship, they have to have roughly the same environment ranks. 

\begin{figure*}
\begin{center}
\includegraphics[totalheight=0.3\textheight,viewport=20 0 870 220,clip]{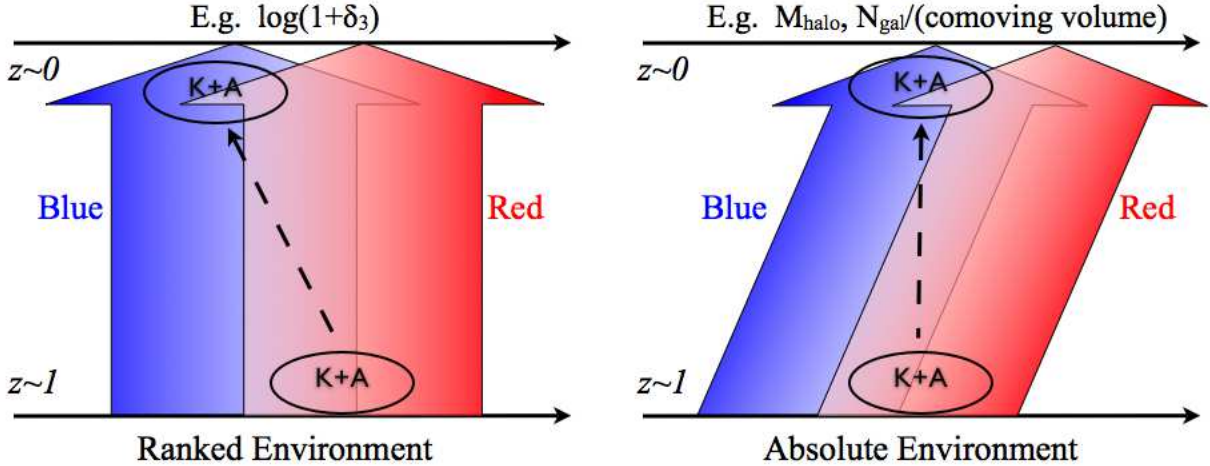}
\caption{These schematic diagrams show the movements of galaxy environments with redshift. The left panel shows the movements in relative environment or ranked environment. The right panel shows the movements in absolute environment (e.g., dark matter halo mass, or the local comoving number density of galaxies). Thick red (blue) arrows represent the movements of red (blue) galaxies. K+A galaxies at low-$z$ are found at lower rank environments than high-$z$ K+As. See \S\ref{sec:discuss} for full discussion. Note, the dashed arrows linking high-$z$ K+As with low-$z$ K+As do not represent a progenitor-descendant relationship: they are completely different populations.
}
\label{fig:env_graph}
\end{center}
\end{figure*}

For the discussion of K+A environment evolution, it is helpful to visualize the environment evolution of all galaxies in a schematic diagram. The left panel of Fig.~\ref{fig:env_graph} shows the movement of galaxies in ranked environment from $z\sim1$ to $z\sim0$. Overall, environment ranks of galaxies do not change significantly over cosmic time. Red galaxies, on average, always rank higher in environment than blue galaxies. Our results show that K+A galaxies at $z\sim1$ are ranked similar to red galaxies, but the equivalent population at $z\sim0$ has much lower ranks and is similar to blue galaxies. This gives us two messages. First, K+As at $z\sim0$ are a very different population from the progenitors of low-$z$ red galaxies. Second, the building-up of the red sequence through the K+A phase is happening in relatively overdense (higher rank) environments at $z\sim1$ but in relatively underdense (lower rank) environments at $z\sim0$. This statement may or may not apply to the overall build-up of the red sequence, depending on whether the K+A phase is the dominant route for the production of a red galaxy, and on the environment evolution of other routes. Nonetheless, our result at $z\sim1$ is consistent with the results from other environment studies. \cite{CooperNC07} and \cite{Gerke07} both showed that between redshifts 0.7 and 1.3 the red sequence was preferentially built up in overdense environments. 

Although high-$z$ K+As have much higher environment ranks than low-$z$ K+As, they may have the same absolute environments. This is because the absolute environments for most galaxies have grown with time. In regions with matter density above the critical density, the density will grow with time, and so will the galaxy number density, if galaxies trace the dark matter distribution on large scales. Most galaxies are in such regions; thus their absolute environments will grow with time. Therefore, the lower environment rank of the K+A population at lower redshift does not necessarily indicate a difference in absolute environment from high-$z$ K+As. This is illustrated in the right panel of Fig.~\ref{fig:env_graph}. The absolute environments of K+As at $z\sim0$ may be the same, or less dense than, the absolute environments of K+As at $z\sim1$; but they cannot be much denser than $z\sim1$. Other than that, we do not have any constraint on this because we only have a relative environment measure.

If we assume K+As at all redshifts are quenched in the same absolute environments and every red galaxy went through this phase, then we can explain the larger environment difference between red galaxies and K+As at low $z$ than that observed at high $z$. The average age of stellar populations in red galaxies is much older on average at $z\sim0.1$ than at $z\sim0.8$ \citep{Gallazzi05, Schiavon06}. A typical low-$z$ red galaxy would probably have quenched a longer amount of time ago, thus its local environment would have more time to grow (e.g., the group it belongs to could accrete more members and/or merge into a cluster) compared to a high-$z$ red galaxy. This will lead to a larger difference in environment with K+As at low-$z$ than high-$z$, as observed: high-$z$ red galaxies still reside in environments similar to those of K+As, maybe only slightly denser; while low-$z$ red galaxies reside in much denser environments than low-$z$ K+As.

This picture not only explains our results with regard to red galaxies, it also has a prediction. Under this scenario, the absolute environments of red galaxies will be correlated with the time passed since quenching, and hence, approximately, the stellar age. If confined to a single redshift epoch, the same statement can be applied to relative environments as well. This will be investigated in a forthcoming paper (Cooper et al. in prep).



\subsection{Quenching Probability Among Blue Galaxies}
The blue galaxy population must include the direct progenitors of future K+A galaxies. Comparing the environment distribution of K+As with that of blue galaxies will reveal how quenching may depend on environment and/or internal properties of galaxies. This dependence also needs to produce the observed evolution in K+A environments with regard to that of blue galaxies. We discuss these topics in this section.


Only galaxies which possessed a significant (by fraction) young stellar population before quenching can appear as a K+A after quenching.  If K+A galaxies are produced from a starburst and then quenching associated with  a merger event, this would be a constraint on the amount of gas available for star formation (as a fraction of stellar mass) before the merger.  Hence, the observed environment distribution of K+A galaxies will be a result of both any environment dependence in the quenching mechanism and any environment dependence in pre-quenching star formation in galaxies.  To further complicate matters, the quenching probability could depend on galaxy properties which correlate with environment, such as stellar mass or SFR, as well as the environment itself.



At $z\sim0.8$, we found that the environment distribution of K+A galaxies is marginally different from that of blue galaxies. However, the significance of the difference is limited by the small sample size. We must wait for future data to confirm this result and improve its significance. In the meantime, we will assume that the difference is robust for the purpose of discussion. 
There are three possible explanations for this difference.


The difference can be explained as a dependence of quenching probability on the local environment. Assuming environment is the only factor determining quenching probability and all blue galaxies satisfy can appear as a K+A after quenching, our results indicate that quenching probability at $z\sim0.8$ is higher in denser environments, but at $z\sim0.1$, it is roughly the same for blue galaxies in any environment (where we here mean absolute, rather than relative, overdensity). The trends at the two redshifts can be consistent only if, by $z\sim0.1$, nearly all blue galaxies have left the low-density regions where quenching is less likely to happen. Whether the overall increase in absolute environment for all blue galaxies is large enough to make this difference awaits simulations to answer.

The second possibility is that star formation quenching does not depend on the local environment, but quenching and the K+A prerequisites depend on certain internal properties of a galaxy, which selects preferentially blue galaxies that are found in overdense environments. For example, if quenching requires a sufficiently high stellar mass (e.g., to have a massive enough central SMBH to provide the feedback) and forming a K+A requires a high pre-quenching SFR, the K+A population will carry the environment footprint of those galaxies satisfying these criteria. At $z\sim0.8$, these galaxies do preferentially reside in overdense environments. As shown by \cite{Cooper06}, the average overdensity on the blue cloud at $z\sim1$ is a strong function of $B$-band luminosity, which correlates strongly with stellar mass, with some sensitivity to the presence of young stars \citep{CooperNW08}. In addition,  \cite{Elbaz07} and \cite{CooperNW08} have both shown that the average SFR increases with galaxy overdensity at $z\sim1$. Therefore, K+As at $z\sim0.8$ will be found in more overdense environments than blue galaxies. 

At $z\sim0.1$, the average environment of galaxies on the blue cloud has no or only weak dependence on $^{0.1}i$-band luminosity \citep{Hogg03,BlantonEH05}. The average SFR is also higher in underdense environments \citep{CooperNW08}. 
As a consequence, if K+A progenitors must be high-SFR blue galaxies, and quenching probability increases modestly with density, we might expect low-redshift poststarburst galaxies to have an environment distribution similar to blue galaxies. 
Therefore, the observed environment distribution of K+As relative to blue galaxies may indirectly be a reflection of the correlations between environment and other galaxy properties. 

This possibility can be distinguished from the first by controlling each galaxy property that has an observable correlation with environment. It is also likely that quenching depends on both environment and internal properties of galaxies. Future surveys yielding much larger sample sizes and higher completeness are required to break the possible degeneracies.

The third possibility is that quenching depends on having a fast and significant increase in local overdensity. Only those galaxies that experience a quick change in environment (e.g., by falling into a cluster) have their star formation quenched. Ram-pressure stripping in clusters is an example of a mechanism that has such dependence on environment. We would expect K+As to be found in rich groups and clusters in this case. This may occur, but it does not explain the field K+As. Additionally, it will be difficult to explain the low-$z$ result, where the K+As have an environment distribution indistinguishable from blue galaxies.

\section{Implications for K+A Quenching Mechanisms}




What does the environment distribution of K+As tell us about their formation? First of all,
if a quenching mechanism has a dependence on environment, the environment we should consider is the absolute environment, not the ranked environment. Therefore, we have less constraining power on this issue as our environment measurement is a relative measure. Furthermore, the fact that the ranked environment distribution of K+As at $z\sim0.8$ differs from that at $z\sim0.1$ does not mean that the quenching mechanisms have to be different. In fact, it is possible that K+As reside in similar absolute environments at both redshifts, and are due to the same quenching mechanism.

A variety of quenching mechanisms that can yield K+A galaxies have been proposed since this population was identified. One class of mechanisms operates only in rich cluster environments, such as ram-pressure stripping \citep[e.g.][]{GunnG72, BekkiC03}, galaxy harassment \citep{Moore96}, and strangulation \citep{Balogh00}. Another popular mechanism is through major merger events \citep{MihosH94, SpringelDM05}. In addition, shock heating in massive halos \citep{BirnboimDN07} may also produce K+A galaxies. Below we discuss the relevance of each mechanism and our constraints on its significance at both $z\sim0$ and $z\sim1$.

\subsection{Cluster-specific mechanisms}\label{sec:mech_cluster}

The first mechanism proposed for K+A formation results from interactions of an infalling galaxy with the gravitational potential or intracluster medium of a rich galaxy cluster. The hot intracluster medium could shock or compress the gas in a galaxy producing a starburst, while also stripping the gas in its outskirts and other low density parts of the ISM, eventually halting star formation \citep{GunnG72, Nulsen82, BothunD86,Balsara94,BekkiC03}

Many examples of galaxies undergoing strong ram pressure stripping have been found in nearby clusters (e.g., Virgo, Coma, A1367); asymmetric radio continuum morphology, HI profile, and/or ionized gas tails \citep[e.g.][]{Irwin87, Vollmer00, Vollmer04, Vollmer05, Cortese07, McConnachie07} have provided evidence for recent stripping events. However, the star formation in these galaxies is not always shut down. In some cases, the specific SFR is enhanced (CGCG 97-073 and 97-079, \citealt{GavazziCC95}) or even raised to starburst levels (235144-260358 in Abell 2667, \citealt{Cortese07}); some objects show signs of moderate SF or have SF rejuvenated from stripped gas falling back (e.g. NGC 4522, which has SF in the inner disk and an extraplanar ring, but a K+A-like outer disk, as found by \citealt{Vollmer00} and \citealt{CrowlK06}; or NGC4848, as found by \citealt{Vollmer01}). In two objects, however, there is no ongoing SF and the systems appear as K+A galaxies (NGC 4569, \citealt{Keel96}, and 131124-012040 in Abell 1689, \citealt{Cortese07}) . All these observations suggest that ram-pressure stripping can make K+A galaxies in some rich groups or clusters, but not in every case. This is expected, as the gas in the inner part of massive disks is difficult to strip away. Studies by \cite{Poggianti04} found that K+A galaxies in Coma are all low-luminosity objects ($M_v > -18.5$), quite different from the high-luminosity K+A's studied in this paper ($M_B< -19.7$). If ram pressure stripping is responsible for both low-$z$ and high-$z$ K+As, it should only be more difficult to strip larger galaxies at higher redshifts as cluster masses were lower then.  Therefore, it is not clear from this evidence that ram-pressure stripping could be the dominant K+A formation mechanism at high $z$. 

Because of the paucity of rich clusters in the DEEP2 volume, we are unable to directly address this question at high-$z$. It is possible that cluster-specific mechanisms can make K+As in clusters, but certainly that is not the dominant mechanism for K+A formation. Regardless, given the presence of K+As in low-density regions at $z\sim1$, and the predominance of K+As at $z\sim0$, cluster-specific mechanisms cannot explain the observed characteristics of the K+A population.




\subsection{Galaxy-Galaxy Mergers}

A second mechanism proposed for K+A formation is galaxy-galaxy mergers. Mergers can trigger brief, intense starbursts in galaxies \citep{MihosH94}. The exhaustion of available gas and feedback from supernovae winds or AGN activity \citep{SpringelDM05} could then halt star formation. Many K+As have been observed to have morphological features characteristic of mergers, such as disturbed morphologies and tidal tails \citep{LiuK95}. \cite{Yang08} recently presented the detailed morphologies of 21 low-$z$ K+A galaxies obtained from HST/ACS and WFPC2 observations. They found that more than half of the K+A galaxies have disturbed morphology and/or tidal features, indicating recent merger events.

Is the environment distribution observed for K+A galaxies consistent with this mechanism? Merger rate probably depends on environment. However, as we mentioned earlier, environment may not be the only factor in determining the probability a galaxy evolves into a K+A through merger. For the final product of a merger event to qualify as a K+A, the galaxies involved has to be relatively gas-rich. The star formation quenching effectiveness after the merger can also depend on the existence and mass of a central SMBH \citep{SpringelDM05}, which in turn could depend on the final stellar mass. 

The most favorable environment for galaxy-galaxy mergers is in poor groups because of the low velocity dispersion in these systems \citep{ZabludoffM98}. However, even poor groups with masses as low as $6\times10^{12} h^{-1}~M_\odot$ in our sample generally have $\log (1+\delta_3) \gtrsim 0$, reflecting their strong clustering \citep{CoilGN06}; and the K+A fraction in these poor groups is lower than among galaxies in isolation. It seems that these facts disfavor the merger scenario. Actually, because only gas-rich mergers is capable of producing K+As, what we should be looking at is the K+A rate relative to gas-rich galaxies only, or blue galaxies. 

Relative to blue galaxies, K+As at $z\sim0.8$ are more frequently found in denser environments. This can be consistent with the merger scenario, although the preference for dense environment could partly be due to a quenching requirement on stellar mass.  At $z\sim0.1$, K+A rate relative to blue galaxies probably varies little with environments. To explain this under the merger scenario, all blue galaxies at $z\sim0.1$ have to have equal probabilities to have a major merger. This is not impossible, but we need simulations to answer this, which is missing in the current literature.

\subsection{Halo Quenching}

A third quenching mechanism that could produce a K+A galaxy, but has not often been mentioned in K+A literature, is through shock heating of the halo gas and the inflowing gas when a halo grows above a certain threshold in mass. The idea dates back to the early work of \cite{ReesO77}, which concluded that halos with $M>10^{12} M_\odot$ would experience a quasi-static contraction phase due to virial shocks. A series of studies by \cite{BirnboimD03} and \cite{DekelB06} using spherical halo gas-accretion simulations, and by \cite{Ocvirk08} using N-body and hydrodynamic simulations, concluded that when a dark matter halo grows above a critical mass ($M_{\rm crit} \sim 10^{12}M_\odot$), shocks will form in these halos, slow down the infalling material and heat it up, thus removing the cold gas supply for star formation. Recent work by \cite{BirnboimDN07} has noted that the whole process actually involves a sequence of quenching {\it and bursting} events due to shock instabilities at early stages. In cases where shocks form relatively late -- at $z<2.5$ -- and the subsequent burst is at $z<1.4$, the final post-burst shock heating is both rapid and effective, and it provides long term suppression of cold gas accretion onto the central disk of the halo. However, the implications of this model for the SFR in the central galaxy are not straightforward. The gas accretion rate on the central disk provides only an upper limit to the SFR; the actual SFR evolution could be more complicated. However, the overall burst-followed-by-quenching trend should remain robust. Therefore, this model provides a mechanism for forming K+A galaxies. If it is the dominant mechanism, K+A galaxies will mostly be found in halos that just crossed the mass threshold. 

It is possible for this mechanism to explain the environment evolution of K+As, since the absolute environment for K+As may not change significantly from high-$z$ to low-$z$. Same halo mass will correspond to different relative environments at different redshifts. This can be tested with a simulation that has the correct clustering strength for both blue and red galaxies, at both $z\sim0.1$ and $z\sim0.8$. We defer this to future work.

\section{Summary and Conclusions}

We have compared the environment distributions of the post-quenching/K+A galaxies with those of red and blue galaxies 
at both $z\sim0.1$ (from SDSS) and $z\sim0.8$ (from DEEP2), using uniformly selected samples defined by an \hb\ EW limit rather than a less-robust \oii\ EW limit. We have found the following facts.

\begin{enumerate}
\item
At high-$z$ ($z\sim0.8$), the environment distribution of the K+A galaxies is indistinguishable from that of the red sequence galaxies, preferring overdense environments, but showing modest difference from that of the blue galaxies.

\item
At low-$z$ ($z\sim0.1$), K+A galaxies are found to inhabit environments dramatically different from those of red galaxies. Instead, local post-quenching systems are observed to have similar environment distribution as blue galaxies, preferring underdense environments. K+As at low redshift have lower rank in the environment distribution of bright galaxies than their high redshift counterparts do.

\item
We do not find any significant dependence on luminosity for the K+A environment distribution, at both $z\sim0.1$ and $z\sim0.8$. 

\item
At $z\sim0.8$, the K+A fraction among DEEP2 group galaxies is consistent with the fraction among field galaxies within 1-$\sigma$.
\end{enumerate}

Based on these facts, we have reached the following conclusions.


\begin{enumerate}

\item 
The quenching of star formation and the build-up of the red sequence through the K+A phase is happening in relatively overdense environments at $z\sim1$ but in relatively underdense environments at $z\sim0$. The absolute environments (i.e., local mass density) for these processes, however, may stay the same with time, as may the quenching mechanisms. This can be true while the relative quenching environment decreases because the mean absolute overdensities of all galaxies has to increase with time due to gravity. 

\item
At $z\sim0.8$, the typical environment for quenching has not changed significantly from the time $L_*$ red galaxies at $z\sim0.8$ were quenched, suggesting K+As at $z\sim0.8$ may be similar to the progenitors of red galaxies found at the same redshift. In contrast, the low-$z$ K+As, having much lower rank in environment distribution than red galaxies, have to be a totally different population from the progenitors of low-$z$ red galaxies. 

\item
At high-$z$, the modest difference between K+As and blue galaxies in environment distribution suggests that either the quenching probability is a function of environment, prefering overdense to underdense environments, or that quenching only happens to a subpopulation of blue galaxies which are preferentially found in overdense environments. 

\item
At low-$z$, K+As have an environment distribution indistinguishable from that of the blue galaxy population. This indicates that quenching probability has only a weak dependence on environment today. 

\item
Although we are unable to directly address the incidence of K+As in clusters due to the paucity of clusters in the DEEP2 volume, the existence of large K+A population in the field at both low-$z$ and high-$z$ indicate that cluster-specific K+A formation mechanisms cannot be the dominant route by which these galaxies are formed. 

\end{enumerate}

Throughout this work, we have also learned a dearly-bought lesson of relying on measurements of K+A fraction as a function of environment. Because of the large uncertainties of environment measurements and the rarity of K+A galaxies, the significance of a K+A fraction vs. environment trend is easily overestimated and can change significantly when bin definitions are altered. More importantly, these trends are very sensitive to the composition of the all-galaxy sample out of which the fraction is defined. It is likely that these issues have led to the widely varying and sometimes contradictory K+A fraction-environment relations found in the literature. Therefore, we warn on blindly making comparisons of these trends between samples defined using different limiting bands, different luminosity cuts, with different selection rates for red and blue galaxies, or at different redshifts. As a result, we recommend comparing the full environment distributions of different samples, using robust, nonparametric statistical tests, rather than focusing on unreliable measurements of K+A fractions.

\section*{Acknowledgments}

RY would like to thank Ann Zabludoff, Christy Tremonti, Yujin Yang, 
Howard Yee, and Erica Ellingson for enlightening discussions. RY acknowledges partial support 
by the Ontario Post-doctoral Fellowship. 
ALC acknowledges support by NASA through Hubble Fellowship grant 
HST-HF-01182.01-A, awarded by the Space Telescope Science Institute, 
which is operated by the AURA Inc. under NASA contract NAS 5-26555.
MCC acknowledges support by NASA through the Spitzer Space Telescope 
Fellowship program.
BFG was supported by the U.S. Department of Energy under contract number
DE-AC02-76SF00515.
The project was supported in part by the NSF grants AST00-71198 and 
AST00-71048. This research made use of the NASA Astrophysics Data System, 
and employed open-source software written and maintained by David Schlegel, 
Douglas Finkbeiner, and others. 

We would like to thank Greg Wirth and all of the Keck Observatory staff 
for their help in the acquisition of the 
Keck/DEIMOS data. We also wish to recognize and acknowledge the highly significant cultural role and reverence that the summit of Mauna Kea has always had within the indigenous Hawaiian community. It is a privilege to be given the opportunity to conduct observations from this mountain.

Funding for the Sloan Digital Sky Survey (SDSS) has been provided 
 by the Alfred P. Sloan Foundation, the Participating Institutions, 
 the National Aeronautics and Space Administration, the National 
 Science Foundation, the U.S. Department of Energy, the Japanese 
 Monbukagakusho, and the Max Planck Society. The SDSS Web site is 
 http://www.sdss.org/.

The SDSS is managed by the Astrophysical Research Consortium (ARC) 
 for the Participating Institutions. The Participating Institutions are 
 The University of Chicago, Fermilab, the Institute for Advanced Study, 
 the Japan Participation Group, The Johns Hopkins University, Los 
 Alamos National Laboratory, the Max-Planck-Institute for Astronomy 
 (MPIA), the Max-Planck-Institute for Astrophysics (MPA), New Mexico 
 State University, University of Pittsburgh, Princeton University, the 
 United States Naval Observatory, and the University of Washington. 

\bibliographystyle{mn2e}
\bibliography{apj-jour,astro_refs}

\begin{thebibliography}{}

\bibitem[\protect\citeauthoryear{{Anderson} \& {Darling}}{{Anderson} \&
  {Darling}}{1954}]{AndersonD54}
{Anderson} T.~W.,  {Darling} D.~A.,  1954, J. Am. Statist. Assoc., 49, 765

\bibitem[\protect\citeauthoryear{{Balogh}, {Miller}, {Nichol}, {Zabludoff} \&
  {Goto}}{{Balogh} et~al.}{2005}]{Balogh05}
{Balogh} M.~L.,  {Miller} C.,  {Nichol} R.,  {Zabludoff} A.,    {Goto} T.,
  2005, \mnras, 360, 587

\bibitem[\protect\citeauthoryear{{Balogh}, {Morris}, {Yee}, {Carlberg} \&
  {Ellingson}}{{Balogh} et~al.}{1999}]{Balogh99}
{Balogh} M.~L.,  {Morris} S.~L.,  {Yee} H.~K.~C.,  {Carlberg} R.~G.,
  {Ellingson} E.,  1999, \apj, 527, 54

\bibitem[\protect\citeauthoryear{{Balogh}, {Navarro} \& {Morris}}{{Balogh}
  et~al.}{2000}]{Balogh00}
{Balogh} M.~L.,  {Navarro} J.~F.,    {Morris} S.~L.,  2000, \apj, 540, 113

\bibitem[\protect\citeauthoryear{{Balsara}, {Livio} \& {O'Dea}}{{Balsara}
  et~al.}{1994}]{Balsara94}
{Balsara} D.,  {Livio} M.,    {O'Dea} C.~P.,  1994, \apj, 437, 83

\bibitem[\protect\citeauthoryear{{Bekki} \& {Couch}}{{Bekki} \&
  {Couch}}{2003}]{BekkiC03}
{Bekki} K.,  {Couch} W.~J.,  2003, \apjl, 596, L13

\bibitem[\protect\citeauthoryear{{Bell} et~al.,}{{Bell}
  et~al.}{2004}]{BellWM04}
{Bell} E.~F.,  et~al., 2004, \apj, 608, 752

\bibitem[\protect\citeauthoryear{{Birnboim} \& {Dekel}}{{Birnboim} \&
  {Dekel}}{2003}]{BirnboimD03}
{Birnboim} Y.,  {Dekel} A.,  2003, \mnras, 345, 349

\bibitem[\protect\citeauthoryear{{Birnboim}, {Dekel} \& {Neistein}}{{Birnboim}
  et~al.}{2007}]{BirnboimDN07}
{Birnboim} Y.,  {Dekel} A.,    {Neistein} E.,  2007, \mnras, 380, 339

\bibitem[\protect\citeauthoryear{{Blake} et~al.,}{{Blake}
  et~al.}{2004}]{BlakePC04}
{Blake} C.,  et~al., 2004, \mnras, 355, 713

\bibitem[\protect\citeauthoryear{{Blanton}}{{Blanton}}{2006}]{Blanton06}
{Blanton} M.~R.,  2006, \apj, 648, 268

\bibitem[\protect\citeauthoryear{{Blanton}, {Eisenstein}, {Hogg}, {Schlegel} \&
  {Brinkmann}}{{Blanton} et~al.}{2005}]{BlantonEH05}
{Blanton} M.~R.,  {Eisenstein} D.,  {Hogg} D.~W.,  {Schlegel} D.~J.,
  {Brinkmann} J.,  2005, \apj, 629, 143

\bibitem[\protect\citeauthoryear{{Blanton} et~al.,}{{Blanton}
  et~al.}{2003}]{BlantonBC03}
{Blanton} M.~R.,  et~al., 2003, \aj, 125, 2348

\bibitem[\protect\citeauthoryear{{Blanton} et~al.,}{{Blanton}
  et~al.}{2005}]{BlantonSS05}
{Blanton} M.~R.,  et~al., 2005, \aj, 129, 2562

\bibitem[\protect\citeauthoryear{{Bothun} \& {Dressler}}{{Bothun} \&
  {Dressler}}{1986}]{BothunD86}
{Bothun} G.~D.,  {Dressler} A.,  1986, \apj, 301, 57

\bibitem[\protect\citeauthoryear{{Brown}, {Dey}, {Jannuzi}, {Brand}, {Benson},
  {Brodwin}, {Croton} \& {Eisenhardt}}{{Brown} et~al.}{2007}]{Brown07}
{Brown} M.~J.~I.,  {Dey} A.,  {Jannuzi} B.~T.,  {Brand} K.,  {Benson} A.~J.,
  {Brodwin} M.,  {Croton} D.~J.,    {Eisenhardt} P.~R.,  2007, \apj, 654, 858

\bibitem[\protect\citeauthoryear{{Bruzual} \& {Charlot}}{{Bruzual} \&
  {Charlot}}{2003}]{BC03}
{Bruzual} G.,  {Charlot} S.,  2003, \mnras, 344, 1000

\bibitem[\protect\citeauthoryear{{Bundy}, {Ellis}, {Conselice}, {Taylor},
  {Cooper}, {Willmer}, {Weiner}, {Coil}, {Noeske} \& {Eisenhardt}}{{Bundy}
  et~al.}{2006}]{Bundy06}
{Bundy} K.,  {Ellis} R.~S.,  {Conselice} C.~J.,  {Taylor} J.~E.,  {Cooper}
  M.~C.,  {Willmer} C.~N.~A.,  {Weiner} B.~J.,  {Coil} A.~L.,  {Noeske} K.~G.,
    {Eisenhardt} P.~R.~M.,  2006, \apj, 651, 120

\bibitem[\protect\citeauthoryear{{Coil}, {Gerke}, {Newman}, {Ma}, {Yan},
  {Cooper}, {Davis}, {Faber}, {Guhathakurta} \& {Koo}}{{Coil}
  et~al.}{2006}]{CoilGN06}
{Coil} A.~L.,  {Gerke} B.~F.,  {Newman} J.~A.,  {Ma} C.-P.,  {Yan} R.,
  {Cooper} M.~C.,  {Davis} M.,  {Faber} S.~M.,  {Guhathakurta} P.,    {Koo}
  D.~C.,  2006, \apj, 638, 668

\bibitem[\protect\citeauthoryear{{Coil}, {Newman}, {Kaiser}, {Davis}, {Ma},
  {Kocevski} \& {Koo}}{{Coil} et~al.}{2004}]{CoilNK04}
{Coil} A.~L.,  {Newman} J.~A.,  {Kaiser} N.,  {Davis} M.,  {Ma} C.-P.,
  {Kocevski} D.~D.,    {Koo} D.~C.,  2004, \apj, 617, 765

\bibitem[\protect\citeauthoryear{{Cooper} et~al.,}{{Cooper}
  et~al.}{2006}]{Cooper06}
{Cooper} M.~C.,  et~al., 2006, \mnras, 370, 198

\bibitem[\protect\citeauthoryear{{Cooper} et~al.,}{{Cooper}
  et~al.}{2007}]{CooperNC07}
{Cooper} M.~C.,  et~al., 2007, \mnras, pp 224--+

\bibitem[\protect\citeauthoryear{{Cooper}, {Newman}, {Madgwick}, {Gerke}, {Yan}
  \& {Davis}}{{Cooper} et~al.}{2005}]{Cooper05}
{Cooper} M.~C.,  {Newman} J.~A.,  {Madgwick} D.~S.,  {Gerke} B.~F.,  {Yan} R.,
    {Davis} M.,  2005, \apj, 634, 833

\bibitem[\protect\citeauthoryear{{Cooper}, {Newman}, {Weiner}, {Yan},
  {Willmer}, {Bundy}, {Coil}, {Conselice}, {Davis}, {Faber}, {Gerke},
  {Guhathakurta}, {Koo} \& {Noeske}}{{Cooper} et~al.}{2008}]{CooperNW08}
{Cooper} M.~C.,  {Newman} J.~A.,  {Weiner} B.~J.,  {Yan} R.,  {Willmer}
  C.~N.~A.,  {Bundy} K.,  {Coil} A.~L.,  {Conselice} C.~J.,  {Davis} M.,
  {Faber} S.~M.,  {Gerke} B.~F.,  {Guhathakurta} P.,  {Koo} D.~C.,    {Noeske}
  K.~G.,  2008, \mnras, 383, 1058

\bibitem[\protect\citeauthoryear{{Cortese}, {Marcillac}, {Richard},
  {Bravo-Alfaro}, {Kneib}, {Rieke}, {Covone}, {Egami}, {Rigby}, {Czoske} \&
  {Davies}}{{Cortese} et~al.}{2007}]{Cortese07}
{Cortese} L.,  {Marcillac} D.,  {Richard} J.,  {Bravo-Alfaro} H.,  {Kneib}
  J.-P.,  {Rieke} G.,  {Covone} G.,  {Egami} E.,  {Rigby} J.,  {Czoske} O.,
  {Davies} J.,  2007, \mnras, 376, 157

\bibitem[\protect\citeauthoryear{{Crowl} \& {Kenney}}{{Crowl} \&
  {Kenney}}{2006}]{CrowlK06}
{Crowl} H.~H.,  {Kenney} J.~D.~P.,  2006, \apjl, 649, L75

\bibitem[\protect\citeauthoryear{{Davis} et~al.,}{{Davis}
  et~al.}{2003}]{Davis03}
{Davis} M.,  et~al., 2003, in {Guhathakurta} P.,  ed., Discoveries and Research
  Prospects from 6- to 10-Meter-Class Telescopes II. Edited by Guhathakurta,
  Puragra. Proceedings of the SPIE, Volume 4834, pp. 161-172 (2003). {Science
  Objectives and Early Results of the DEEP2 Redshift Survey}.
pp 161--172

\bibitem[\protect\citeauthoryear{{Davis} et~al.,}{{Davis}
  et~al.}{2007}]{Davis07}
{Davis} M.,  et~al., 2007, \apjl, 660, L1

\bibitem[\protect\citeauthoryear{{Dekel} \& {Birnboim}}{{Dekel} \&
  {Birnboim}}{2006}]{DekelB06}
{Dekel} A.,  {Birnboim} Y.,  2006, \mnras, 368, 2

\bibitem[\protect\citeauthoryear{{Dressler} \& {Gunn}}{{Dressler} \&
  {Gunn}}{1983}]{DresslerG83}
{Dressler} A.,  {Gunn} J.~E.,  1983, \apj, 270, 7

\bibitem[\protect\citeauthoryear{{Dressler}, {Smail}, {Poggianti}, {Butcher},
  {Couch}, {Ellis} \& {Oemler}}{{Dressler} et~al.}{1999}]{Dressler99}
{Dressler} A.,  {Smail} I.,  {Poggianti} B.~M.,  {Butcher} H.,  {Couch} W.~J.,
  {Ellis} R.~S.,    {Oemler} A.~J.,  1999, \apjs, 122, 51

\bibitem[\protect\citeauthoryear{{Elbaz}, {Daddi}, {Le Borgne}, {Dickinson},
  {Alexander}, {Chary}, {Starck}, {Brandt}, {Kitzbichler}, {MacDonald},
  {Nonino}, {Popesso}, {Stern} \& {Vanzella}}{{Elbaz} et~al.}{2007}]{Elbaz07}
{Elbaz} D.,  {Daddi} E.,  {Le Borgne} D.,  {Dickinson} M.,  {Alexander} D.~M.,
  {Chary} R.-R.,  {Starck} J.-L.,  {Brandt} W.~N.,  {Kitzbichler} M.,
  {MacDonald} E.,  {Nonino} M.,  {Popesso} P.,  {Stern} D.,    {Vanzella} E.,
  2007, \aap, 468, 33

\bibitem[\protect\citeauthoryear{{Faber} et~al.,}{{Faber}
  et~al.}{2007}]{Faber07}
{Faber} S.~M.,  et~al., 2007, \apj, 665, 265

\bibitem[\protect\citeauthoryear{{Fukugita}, {Ichikawa}, {Gunn}, {Doi},
  {Shimasaku} \& {Schneider}}{{Fukugita} et~al.}{1996}]{Fukugita96}
{Fukugita} M.,  {Ichikawa} T.,  {Gunn} J.~E.,  {Doi} M.,  {Shimasaku} K.,
  {Schneider} D.~P.,  1996, \aj, 111, 1748

\bibitem[\protect\citeauthoryear{{Gallazzi}, {Charlot}, {Brinchmann}, {White}
  \& {Tremonti}}{{Gallazzi} et~al.}{2005}]{Gallazzi05}
{Gallazzi} A.,  {Charlot} S.,  {Brinchmann} J.,  {White} S.~D.~M.,
  {Tremonti} C.~A.,  2005, \mnras, 362, 41

\bibitem[\protect\citeauthoryear{{Gavazzi}, {Contursi}, {Carrasco}, {Boselli},
  {Kennicutt}, {Scodeggio} \& {Jaffe}}{{Gavazzi} et~al.}{1995}]{GavazziCC95}
{Gavazzi} G.,  {Contursi} A.,  {Carrasco} L.,  {Boselli} A.,  {Kennicutt} R.,
  {Scodeggio} M.,    {Jaffe} W.,  1995, \aap, 304, 325

\bibitem[\protect\citeauthoryear{{Gerke}, {Newman}, {Davis}, {Marinoni}, {Yan},
  {Coil}, {Conroy}, {Cooper}, {Faber}, {Finkbeiner}, {Guhathakurta}, {Kaiser},
  {Koo}, {Phillips}, {Weiner} \& {Willmer}}{{Gerke} et~al.}{2005}]{Gerke05}
{Gerke} B.~F.,  {Newman} J.~A.,  {Davis} M.,  {Marinoni} C.,  {Yan} R.,  {Coil}
  A.~L.,  {Conroy} C.,  {Cooper} M.~C.,  {Faber} S.~M.,  {Finkbeiner} D.~P.,
  {Guhathakurta} P.,  {Kaiser} N.,  {Koo} D.~C.,  {Phillips} A.~C.,  {Weiner}
  B.~J.,    {Willmer} C.~N.~A.,  2005, \apj, 625, 6

\bibitem[\protect\citeauthoryear{{Gerke}, {Newman}, {Faber}, {Cooper},
  {Croton}, {Davis}, {Willmer}, {Yan}, {Coil}, {Guhathakurta}, {Koo} \&
  {Weiner}}{{Gerke} et~al.}{2007}]{Gerke07}
{Gerke} B.~F.,  {Newman} J.~A.,  {Faber} S.~M.,  {Cooper} M.~C.,  {Croton}
  D.~J.,  {Davis} M.,  {Willmer} C.~N.~A.,  {Yan} R.,  {Coil} A.~L.,
  {Guhathakurta} P.,  {Koo} D.~C.,    {Weiner} B.~J.,  2007, \mnras, 376, 1425

\bibitem[\protect\citeauthoryear{{Goto}}{{Goto}}{2005}]{Goto05}
{Goto} T.,  2005, \mnras, 357, 937

\bibitem[\protect\citeauthoryear{{Goto}}{{Goto}}{2007}]{Goto07}
{Goto} T.,  2007, \mnras, 377, 1222

\bibitem[\protect\citeauthoryear{{Goto} et~al.,}{{Goto}  et~al.}{2003}]{Goto03}
{Goto} T.,  et~al., 2003, \pasj, 55, 771

\bibitem[\protect\citeauthoryear{{Gunn} \& {Gott}}{{Gunn} \&
  {Gott}}{1972}]{GunnG72}
{Gunn} J.~E.,  {Gott} J.~R.~I.,  1972, \apj, 176, 1

\bibitem[\protect\citeauthoryear{{Ho}, {Filippenko} \& {Sargent}}{{Ho}
  et~al.}{1997}]{HoFS97V}
{Ho} L.~C.,  {Filippenko} A.~V.,    {Sargent} W.~L.~W.,  1997, \apj, 487, 568

\bibitem[\protect\citeauthoryear{{Ho}}{{Ho}}{2004}]{Ho04}
{Ho} L.~C.~W.,  2004, in Coevolution of Black Holes and Galaxies {Black Hole
  Demography from Nearby Active Galactic Nuclei}.
pp 293--+

\bibitem[\protect\citeauthoryear{{Hogg} et~al.,}{{Hogg}  et~al.}{2003}]{Hogg03}
{Hogg} D.~W.,  et~al., 2003, \apjl, 585, L5

\bibitem[\protect\citeauthoryear{{Hogg}, {Masjedi}, {Berlind}, {Blanton},
  {Quintero} \& {Brinkmann}}{{Hogg} et~al.}{2006}]{Hogg06}
{Hogg} D.~W.,  {Masjedi} M.,  {Berlind} A.~A.,  {Blanton} M.~R.,  {Quintero}
  A.~D.,    {Brinkmann} J.,  2006, \apj, 650, 763

\bibitem[\protect\citeauthoryear{{Irwin}, {Seaquist}, {Taylor} \&
  {Duric}}{{Irwin} et~al.}{1987}]{Irwin87}
{Irwin} J.~A.,  {Seaquist} E.~R.,  {Taylor} A.~R.,    {Duric} N.,  1987, \apjl,
  313, L91

\bibitem[\protect\citeauthoryear{{Keel}}{{Keel}}{1996}]{Keel96}
{Keel} W.~C.,  1996, \pasp, 108, 917

\bibitem[\protect\citeauthoryear{{Le Borgne}, {Abraham}, {Daniel}, {McCarthy},
  {Glazebrook}, {Savaglio}, {Crampton}, {Juneau}, {Carlberg}, {Chen}, {Marzke},
  {Roth}, {J{\o}rgensen} \& {Murowinski}}{{Le Borgne}
  et~al.}{2006}]{LeBorgne06}
{Le Borgne} D.,  {Abraham} R.,  {Daniel} K.,  {McCarthy} P.~J.,  {Glazebrook}
  K.,  {Savaglio} S.,  {Crampton} D.,  {Juneau} S.,  {Carlberg} R.~G.,  {Chen}
  H.-W.,  {Marzke} R.~O.,  {Roth} K.,  {J{\o}rgensen} I.,    {Murowinski} R.,
  2006, \apj, 642, 48

\bibitem[\protect\citeauthoryear{{Le Borgne}, {Rocca-Volmerange}, {Prugniel},
  {Lan{\c c}on}, {Fioc} \& {Soubiran}}{{Le Borgne} et~al.}{2004}]{LeBorgne04}
{Le Borgne} D.,  {Rocca-Volmerange} B.,  {Prugniel} P.,  {Lan{\c c}on} A.,
  {Fioc} M.,    {Soubiran} C.,  2004, \aap, 425, 881

\bibitem[\protect\citeauthoryear{{Liu} \& {Kennicutt} Jr.}{{Liu} \&
  {Kennicutt}}{1995}]{LiuK95}
{Liu} C.~T.,  {Kennicutt} Jr. R.~C.,  1995, \apj, 450, 547

\bibitem[\protect\citeauthoryear{{Mann} \& {Whitney}}{{Mann} \&
  {Whitney}}{1947}]{MannW47}
{Mann} H.~B.,  {Whitney} D.~R.,  1947, The Annals of Mathematical Statistics,
  18, 50

\bibitem[\protect\citeauthoryear{{Marinoni}, {Davis}, {Newman} \&
  {Coil}}{{Marinoni} et~al.}{2002}]{Marinoni02}
{Marinoni} C.,  {Davis} M.,  {Newman} J.~A.,    {Coil} A.~L.,  2002, \apj, 580,
  122

\bibitem[\protect\citeauthoryear{{McConnachie}, {Venn}, {Irwin}, {Young} \&
  {Geehan}}{{McConnachie} et~al.}{2007}]{McConnachie07}
{McConnachie} A.~W.,  {Venn} K.~A.,  {Irwin} M.~J.,  {Young} L.~M.,    {Geehan}
  J.~J.,  2007, \apjl, 671, L33

\bibitem[\protect\citeauthoryear{{Mihos} \& {Hernquist}}{{Mihos} \&
  {Hernquist}}{1994}]{MihosH94}
{Mihos} J.~C.,  {Hernquist} L.,  1994, \apjl, 431, L9

\bibitem[\protect\citeauthoryear{{Moore}, {Katz}, {Lake}, {Dressler} \&
  {Oemler}}{{Moore} et~al.}{1996}]{Moore96}
{Moore} B.,  {Katz} N.,  {Lake} G.,  {Dressler} A.,    {Oemler} A.,  1996,
  \nat, 379, 613

\bibitem[\protect\citeauthoryear{{Noeske}, {Faber}, {Weiner}, {Koo}, {Primack},
  {Dekel}, {Papovich}, {Conselice}, {Le Floc'h}, {Rieke}, {Coil}, {Lotz},
  {Somerville} \& {Bundy}}{{Noeske} et~al.}{2007}]{NoeskeFW07}
{Noeske} K.~G.,  {Faber} S.~M.,  {Weiner} B.~J.,  {Koo} D.~C.,  {Primack}
  J.~R.,  {Dekel} A.,  {Papovich} C.,  {Conselice} C.~J.,  {Le Floc'h} E.,
  {Rieke} G.~H.,  {Coil} A.~L.,  {Lotz} J.~M.,  {Somerville} R.~S.,    {Bundy}
  K.,  2007, \apjl, 660, L47

\bibitem[\protect\citeauthoryear{{Norton}, {Gebhardt}, {Zabludoff} \&
  {Zaritsky}}{{Norton} et~al.}{2001}]{Norton01}
{Norton} S.~A.,  {Gebhardt} K.,  {Zabludoff} A.~I.,    {Zaritsky} D.,  2001,
  \apj, 557, 150

\bibitem[\protect\citeauthoryear{{Nulsen}}{{Nulsen}}{1982}]{Nulsen82}
{Nulsen} P.~E.~J.,  1982, \mnras, 198, 1007

\bibitem[\protect\citeauthoryear{{Ocvirk}, {Pichon} \& {Teyssier}}{{Ocvirk}
  et~al.}{2008}]{Ocvirk08}
{Ocvirk} P.,  {Pichon} C.,    {Teyssier} R.,  2008, \mnras, 390, 1326

\bibitem[\protect\citeauthoryear{{Oke} \& {Gunn}}{{Oke} \&
  {Gunn}}{1983}]{OkeG83}
{Oke} J.~B.,  {Gunn} J.~E.,  1983, \apj, 266, 713

\bibitem[\protect\citeauthoryear{{Pettitt}}{{Pettitt}}{1976}]{Pettitt76}
{Pettitt} A.~N.,  1976, Biometrika, 63, 161

\bibitem[\protect\citeauthoryear{{Poggianti}, {Bridges}, {Komiyama}, {Yagi},
  {Carter}, {Mobasher}, {Okamura} \& {Kashikawa}}{{Poggianti}
  et~al.}{2004}]{Poggianti04}
{Poggianti} B.~M.,  {Bridges} T.~J.,  {Komiyama} Y.,  {Yagi} M.,  {Carter} D.,
  {Mobasher} B.,  {Okamura} S.,    {Kashikawa} N.,  2004, \apj, 601, 197

\bibitem[\protect\citeauthoryear{{Poggianti} et~al.,}{{Poggianti}
  et~al.}{2009}]{Poggianti09}
{Poggianti} B.~M.,  et~al., 2009, \apj, 693, 112

\bibitem[\protect\citeauthoryear{{Poggianti}, {Smail}, {Dressler}, {Couch},
  {Barger}, {Butcher}, {Ellis} \& {Oemler}}{{Poggianti}
  et~al.}{1999}]{Poggianti99}
{Poggianti} B.~M.,  {Smail} I.,  {Dressler} A.,  {Couch} W.~J.,  {Barger}
  A.~J.,  {Butcher} H.,  {Ellis} R.~S.,    {Oemler} A.~J.,  1999, \apj, 518,
  576

\bibitem[\protect\citeauthoryear{{Quintero} et~al.,}{{Quintero}
  et~al.}{2004}]{Quintero04}
{Quintero} A.~D.,  et~al., 2004, \apj, 602, 190

\bibitem[\protect\citeauthoryear{{Rees} \& {Ostriker}}{{Rees} \&
  {Ostriker}}{1977}]{ReesO77}
{Rees} M.~J.,  {Ostriker} J.~P.,  1977, \mnras, 179, 541

\bibitem[\protect\citeauthoryear{{Schiavon}, {Faber}, {Konidaris}, {Graves},
  {Willmer}, {Weiner}, {Coil}, {Cooper}, {Davis}, {Harker}, {Koo}, {Newman} \&
  {Yan}}{{Schiavon} et~al.}{2006}]{Schiavon06}
{Schiavon} R.~P.,  {Faber} S.~M.,  {Konidaris} N.,  {Graves} G.,  {Willmer}
  C.~N.~A.,  {Weiner} B.~J.,  {Coil} A.~L.,  {Cooper} M.~C.,  {Davis} M.,
  {Harker} J.,  {Koo} D.~C.,  {Newman} J.~A.,    {Yan} R.,  2006, \apjl, 651,
  L93

\bibitem[\protect\citeauthoryear{{Sinclair} \& {Spurr}}{{Sinclair} \&
  {Spurr}}{1988}]{SinclairS88}
{Sinclair} C.~D.,  {Spurr} B.~D.,  1988, J. Am. Statist. Assoc., 83, 1190

\bibitem[\protect\citeauthoryear{{Springel}, {Di Matteo} \&
  {Hernquist}}{{Springel} et~al.}{2005}]{SpringelDM05}
{Springel} V.,  {Di Matteo} T.,    {Hernquist} L.,  2005, \mnras, 361, 776

\bibitem[\protect\citeauthoryear{{Stoughton} et~al.,}{{Stoughton}
  et~al.}{2002}]{Stoughton02}
{Stoughton} C.,  et~al., 2002, \aj, 123, 485

\bibitem[\protect\citeauthoryear{{Tran}, {Franx}, {Illingworth}, {Kelson} \&
  {van Dokkum}}{{Tran} et~al.}{2003}]{TranFI03}
{Tran} K.-V.~H.,  {Franx} M.,  {Illingworth} G.,  {Kelson} D.~D.,    {van
  Dokkum} P.,  2003, \apj, 599, 865

\bibitem[\protect\citeauthoryear{{Tran}, {Franx}, {Illingworth}, {van Dokkum},
  {Kelson} \& {Magee}}{{Tran} et~al.}{2004}]{TranFI04}
{Tran} K.-V.~H.,  {Franx} M.,  {Illingworth} G.~D.,  {van Dokkum} P.,  {Kelson}
  D.~D.,    {Magee} D.,  2004, \apj, 609, 683

\bibitem[\protect\citeauthoryear{{Vollmer}, {Balkowski}, {Cayatte}, {van Driel}
  \& {Huchtmeier}}{{Vollmer} et~al.}{2004}]{Vollmer04}
{Vollmer} B.,  {Balkowski} C.,  {Cayatte} V.,  {van Driel} W.,    {Huchtmeier}
  W.,  2004, \aap, 419, 35

\bibitem[\protect\citeauthoryear{{Vollmer}, {Braine}, {Balkowski}, {Cayatte} \&
  {Duschl}}{{Vollmer} et~al.}{2001}]{Vollmer01}
{Vollmer} B.,  {Braine} J.,  {Balkowski} C.,  {Cayatte} V.,    {Duschl} W.~J.,
  2001, \aap, 374, 824

\bibitem[\protect\citeauthoryear{{Vollmer}, {Huchtmeier} \& {van
  Driel}}{{Vollmer} et~al.}{2005}]{Vollmer05}
{Vollmer} B.,  {Huchtmeier} W.,    {van Driel} W.,  2005, \aap, 439, 921

\bibitem[\protect\citeauthoryear{{Vollmer}, {Marcelin}, {Amram}, {Balkowski},
  {Cayatte} \& {Garrido}}{{Vollmer} et~al.}{2000}]{Vollmer00}
{Vollmer} B.,  {Marcelin} M.,  {Amram} P.,  {Balkowski} C.,  {Cayatte} V.,
  {Garrido} O.,  2000, \aap, 364, 532

\bibitem[\protect\citeauthoryear{{White} et~al.,}{{White}
  et~al.}{2005}]{WhiteCS05}
{White} S.~D.~M.,  et~al., 2005, \aap, 444, 365

\bibitem[\protect\citeauthoryear{{Wild}, {Walcher}, {Johansson}, {Tresse},
  {Charlot}, {Pollo}, {Le F{\`e}vre} \& {de Ravel}}{{Wild}
  et~al.}{2009}]{Wild09}
{Wild} V.,  {Walcher} C.~J.,  {Johansson} P.~H.,  {Tresse} L.,  {Charlot} S.,
  {Pollo} A.,  {Le F{\`e}vre} O.,    {de Ravel} L.,  2009, \mnras, 395, 144

\bibitem[\protect\citeauthoryear{{Willmer} et~al.,}{{Willmer}
  et~al.}{2006}]{Willmer06}
{Willmer} C.~N.~A.,  et~al., 2006, \apj, 647, 853

\bibitem[\protect\citeauthoryear{{Yan}, {Newman}, {Faber}, {Konidaris}, {Koo}
  \& {Davis}}{{Yan} et~al.}{2006}]{Yan06}
{Yan} R.,  {Newman} J.~A.,  {Faber} S.~M.,  {Konidaris} N.,  {Koo} D.,
  {Davis} M.,  2006, \apj, 648, 281

\bibitem[\protect\citeauthoryear{{Yang}, {Zabludoff}, {Zaritsky} \&
  {Mihos}}{{Yang} et~al.}{2008}]{Yang08}
{Yang} Y.,  {Zabludoff} A.~I.,  {Zaritsky} D.,    {Mihos} J.~C.,  2008, \apj,
  688, 945

\bibitem[\protect\citeauthoryear{{Yee}, {Gladders}, {Gilbank}, {Majumdar},
  {Hoekstra} \& {Ellingson}}{{Yee} et~al.}{2007}]{Yee07}
{Yee} H.~K.~C.,  {Gladders} M.~D.,  {Gilbank} D.~G.,  {Majumdar} S.,
  {Hoekstra} H.,    {Ellingson} E.,  2007, in {Metcalfe} N.,  {Shanks} T.,
  eds, Cosmic Frontiers Vol.~379 of Astronomical Society of the Pacific
  Conference Series, {The Red-Sequence Cluster Surveys}.
pp 103--+

\bibitem[\protect\citeauthoryear{{York} et~al.,}{{York}  et~al.}{2000}]{York00}
{York} D.~G.,  et~al., 2000, \aj, 120, 1579

\bibitem[\protect\citeauthoryear{{Zabludoff} \& {Mulchaey}}{{Zabludoff} \&
  {Mulchaey}}{1998}]{ZabludoffM98}
{Zabludoff} A.~I.,  {Mulchaey} J.~S.,  1998, \apj, 496, 39

\bibitem[\protect\citeauthoryear{{Zabludoff}, {Zaritsky}, {Lin}, {Tucker},
  {Hashimoto}, {Shectman}, {Oemler} \& {Kirshner}}{{Zabludoff}
  et~al.}{1996}]{Zabludoff96}
{Zabludoff} A.~I.,  {Zaritsky} D.,  {Lin} H.,  {Tucker} D.,  {Hashimoto} Y.,
  {Shectman} S.~A.,  {Oemler} A.,    {Kirshner} R.~P.,  1996, \apj, 466, 104

\end{thebibliography}

\end{document}